    \documentclass[5p]{elsarticle}  
    
    \usepackage{lineno}
    \usepackage{amsmath}
    \usepackage{siunitx}
    \modulolinenumbers[5]
\usepackage[utf8]{inputenc}
\usepackage[colorlinks,citecolor=blue,linkcolor=blue,urlcolor=blue]{hyperref}
\usepackage{graphicx}
\usepackage{dcolumn}
\usepackage{bm}
\usepackage{color}
\usepackage{amsfonts}
\usepackage{txfonts}
\usepackage{multirow}
\usepackage[english]{babel}
\usepackage{titlesec}
\usepackage{outlines}
\usepackage{comment}
\usepackage{upgreek}
\usepackage{doi}
\usepackage{tablefootnote}

\begin{document}

\pagestyle{myheadings}
\markright{Revised Version}

\begin{frontmatter}

\title{Key points in the determination of the interfacial Dzyaloshinskii-Moriya interaction from asymmetric bubble domain expansion}

\author[1]{A.~Magni}\corref{correspondingauthor}
\cortext[correspondingauthor]{Corresponding author, a.magni@inrim.it}
\author[6]{G.~Carlotti}
\author[1]{A.~Casiraghi}
\author[4]{E.~Darwin}
\author[1]{G.~Durin}

\author [11]{L.~Herrera Diez}
\author[4]{B.~J.~Hickey}
\author[4]{A.~Huxtable}
\author[2]{C.Y.~Hwang}
\author[3]{G.~Jakob}
\author[2]{C.~Kim}
\author[3]{M.~Kläui}
\author[12]{J.~Langer}
\author[4]{C.~H.~Marrows}

\author[7,7a]{H.~T.~Nembach}
\author[11,12]{D.~Ravelosona}

\author[7,8a]{G.~A.~Riley}

\author[7]{J.~M.~Shaw}

\author[10]{V.~Sokalski}
\author[5]{S.~Tacchi}

\author[1]{M.~Kuepferling}

\address[1]{Istituto Nazionale di Ricerca Metrologica, Torino, Italy}
\address[2]{Korea Research Institute of Standards and Science, Yuseong-Gu Daejeon, Republic of Korea}
\address[3]{Institute of Physics, Johannes Gutenberg-University, Mainz, Germany}
\address[4]{School of Physics and Astronomy, University of Leeds, Leeds LS2 9JT, United Kingdom}
\address[5]{CNR, Istituto Officina dei Materiali - Perugia, c/o Dipartimento di Fisica e Geologia\\Università Perugia, Italy}
\address[6]{Dipartimento di Fisica e Geologia, Università di Perugia, Italy}
\address[7]{NIST, National Institute of Standards and Technology, Boulder, Colorado, USA}
\address[7a]{Department of Physics, University of Colorado, USA}
\address[8a]{Center for Memory and Recording Research, University of California-San Diego, La Jolla, USA}
\address[10]{Department of Materials Science and Engineering, Carnegie Mellon University, Pittsburgh, Pennsylvania, USA}
\address[11]{Centre de Nanosciences et de Nanotechnologies, CNRS, Universit\'e Paris-Saclay, 91120 Palaiseau, France.}
\address[12]{Spin-Ion Technologies, C2N, Palaiseau, France}

\date{\today}
\begin{abstract}
Different models have been used to evaluate the interfacial Dzyaloshinskii-Moriya interaction (DMI) from the asymmetric bubble expansion method using magneto-optics.
Here we investigate the most promising candidates over a range of different magnetic multilayers with perpendicular anisotropy.
Models based on the standard creep hypothesis are not able to reproduce the domain wall (DW) velocity profile when the DW roughness is high. Our results demonstrate that the DW roughness and the interface roughness of the sample layers are correlated. Furthermore, we give guidance on how to obtain reliable results for the DMI value with this popular method.
A comparison of the results with Brillouin light scattering (BLS) measurements on the same samples shows that the BLS approach often results in higher measured values of DMI.

\end{abstract}


\end{frontmatter}


\section{Introduction}

One of the effects of the Dzyaloshinskii-Moriya interaction  \cite{DZY-57,MOR-60} (DMI) in non-centrosymmetric bulk magnetic systems is the stabilization of chiral magnetic structures, such as helical magnetization regions and skyrmions \cite{Na-13,GAR-17, EV-18}. This asymmetric exchange interaction, which favors a non-collinear alignment of neighboring spins, was recently rediscovered in heterostructures of potential technological importance, consisting of ferromagnet/heavy metal (FM/HM) thin films \cite{BOD-07,ZAK-10, MOO-13, JE-13}. Such heterostructures, which often present a perpendicular magnetic anisotropy (PMA), are promising for applications as novel magnetic memory, sensors, or logic elements and devices \cite{PAR-15, DIE-20}. The DMI in these heterostructures originates at the interface between the FM layer and the HM layer from spin-orbit coupling and is non-zero due to the symmetry-breaking caused by the presence of the interface. Therefore it is also called interfacial DMI.

In the presence of PMA in a ferromagnetic layer, different domain structures are possible - stripes, bubbles, and cellular domains \cite{CAP-71}. Acting like an effective local in-plane magnetic field, the DMI changes the nature of the DWs of these structures, with Néel walls energetically favored over Bloch walls for strong DMI \cite{BOD-07,CHE-13}. The presence of DMI also alters the dynamic regime, preventing the transition from Néel to Bloch DW up to high fields. Therefore, the Walker breakdown field is remarkably increased \cite{THI-12}, allowing high DW velocities, such that DWs might be efficiently used as information carriers in storage and logic devices  \cite{YA-15}, driven by currents or by magnetic fields \cite{AB-21,LU-20}. This motivated a strong interest in the optimization of the DMI, and the accurate evaluation of its strength.

A measure for the stability of chiral magnetic structures and the strength of the DMI is the related energy coefficient, $\boldsymbol{D}$ \cite{MOR-60}. In general $\boldsymbol{D}$ is a tensor composed of Lifshitz invariants \cite{DZY-64,BOG-89a} that relates the anisotropic exchange energy to the local magnetization. In certain symmetries (as in the case for the interfacial DMI investigated here), $\boldsymbol{D}$ can be reduced to a single scalar DMI value $D$, which is dependent on material and interface quality. 

Different techniques exist to determine the value of $D$, but significant discrepancies are often found when comparing similar thin film stacks measured by different techniques. In fact, each method may require a specific sample preparation, and not every measurement method can perform optimally over a range of sample thickness and roughness. Furthermore, evaluation of the data is often complex and requires a knowledge of material properties with high accuracy, hence contradictory values can be found in the literature \cite{KIM-19a, Kuepferling2021, LOC-17}. 


In this paper, we concentrate on a DW velocity-based measurement method of the $D$ value: the asymmetric expansion of a bubble domain in presence of DMI and an in-plane applied magnetic field \cite{JE-13,HRA-14}. This method, being relatively easily available in many laboratories, since it requires no special sample preparation and only standard magneto-optical equipment, might become a standard technique for DMI measurements. We analyse here the advantages and limitations of the technique with the goal to provide recommendations for a good practice. We follow the whole process, starting from the measurement down to the data processing and the application of a well-defined DMI model. In particular, we review different theoretical models that are used in the literature to evaluate the measured data with the goal to define the categories of sample and measurement conditions to which they are applicable. We emphasize the key points needed to achieve a reliable determination of the DMI value, reducing systematic measurement errors and problems of repeatability. Furthermore, we consider results of an international ``round robin'' (RR) comparison, where identical samples were measured in different laboratories. We therefore verify the achievable measurement reliability by comparison with the most popular method based on spin waves: Brillouin light scattering (BLS).



\section{Samples}

In this work, different sample classes were selected in order to investigate a wide variety of bubble domain structures. Two classical material combinations, Pt/Co and Pt/Co/Pt multilayers at different thicknesses, were produced; additionally we investigated Pt/Co/Ir structures, since in the past controversial results on the DMI sign of Ir were published \cite{YAM-16, YAN-15, LAU-18} (Tab.~\ref{table:Co samples}).

To investigate another FM material with slightly different magnetic properties, we analyzed W/FeCoB/MgO and Pt/CoFeB/MgO systems, single and multiple FM layers (annealed or not), with several repetitions of the HM/FM bilayer stack. 
Since the interface roughness is a critical parameter for DMI it was decided to vary this parameter as well, by tuning the Pt sputtering power (Tab.~\ref{table:CoFeB WPt samples}). RF sputter deposition of Pt at higher power leads to a lower surface roughness, with the presence of columnar growth and a higher packing fraction of Pt \cite{SLA-09}, which was demonstrated on 100~nm thick films. The samples shown in Table \ref{table:Co samples} and \ref{table:CoFeB WPt samples} were prepared for an international interlaboratory comparison of the DMI value measurement \cite{TOPS}.

\begin{table}
\caption{Co samples with Pt and/or Ir heavy metal layer prepared at the University of Leeds.}
\label{table:Co samples}
\small
\begin{tabular}{c|c|c|c}
\hline
\hline
\textbf{Sample}& \textbf{FM layer} & \textbf{Bottom layer} & \textbf{Top layer}\\
\hline
& \textbf{(nm)} & \textbf{(nm)} & \textbf{(nm)}\\
\hline
\hline
a1& \multirow{5}{*}{\centering Co(0.8)} & \multirow{5}{*}{\centering Ta(5)/Pt(3)} & Pt(3)/Ta(3)\\
a2& &&Pt(1)/Ta(3)\\
a3& &&Ir(3)/Ta(3)\\
a4& &&Ir(1)/Ta(3)\\
a5& &&Ta(3)\\
\hline
\hline
\end{tabular}
\end{table}

\begin{table}
\caption{CoFeB-based samples with W (first batch) or Pt (second batch) heavy metal layer prepared at University of Mainz. All samples are prepared on Si substrate (native oxide) with an adhesion layer of 5.7~nm of Ta. In the second batch the Pt sputter power (SP) is varied in order to have different interface roughness. Annealing (\emph{ann.}) was performed at 400$^\circ$C for one hour.}
\label{table:CoFeB WPt samples}
\small
\begin{tabular}{c|c|c|c|c|c}
\hline
\hline
\textbf{\scriptsize{sample}}& \textbf{\scriptsize{FM layer}} & \textbf{\scriptsize{bottom layer}} & \textbf{\scriptsize{top layer}}&\textbf{\scriptsize{ann.}} &\textbf{\scriptsize{SP}}\\
\hline
& \scriptsize{(nm)} & \scriptsize{(nm)} & \scriptsize{(nm)}  & &\scriptsize{(W)}\\
\hline
\hline
756a &  \multirow{2}{*}{\centering\scriptsize{Co$_{20}$Fe$_{60}$B$_{20}$(0.6)}} & \multirow{2}{*}{\centering\scriptsize{W(5)}} & \multirow{2}{*}{\centering\scriptsize{MgO(2)/Ta(5)} }& \scriptsize{yes}&\multirow{2}{*}{\centering\scriptsize{200}}\\
756b &   &  & & \scriptsize{no}&\\
\hline
758a & \multirow{3}{*}{\centering \scriptsize{Co$_{60}$Fe$_{20}$B$_{20}$(0.8)}} & \multirow{3}{*}{\centering{\scriptsize{Pt(3.4)}}}& \multirow{3}{*}{\centering{\scriptsize{MgO(1.4)/Ta(5)}}}&\multirow{3}{*}{\centering{\scriptsize{yes}}} &\scriptsize{200}\\
760a &  & & && \scriptsize{700}\\
762a & & & & &\scriptsize{1200}\\
\hline
759a & \multicolumn{3}{|c|}{\scriptsize{[Pt(3.4)/Co$_{60}$Fe$_{20}$B$_{20}$(0.8)/MgO(1.4)]x5/Ta(5)}} &\scriptsize{yes}& \scriptsize{200}\\
763a & \multicolumn{3}{|c|}{\scriptsize{[Pt(3.4)/Co$_{60}$Fe$_{20}$B$_{20}$(0.8)/MgO(1.4)]x5/Ta(5)}}  &\scriptsize{yes}& \scriptsize{1200}\\
\hline
\hline
\end{tabular}
\end{table}

Last, Ta/CoFeB/MgO films were produced, with a range of He$^+$ ion fluences  (Tab.~\ref{table:implanted samples}). Irradiation increases the DMI strength and reduces the saturation magnetization, as a consequence of the effects of ion irradiation on the bottom and top CoFeB interfaces \cite{HER-15}.

All thicknesses reported in the Tab.~\ref{table:Co samples} to \ref{table:implanted samples} (the numbers in brackets) are nominal thicknesses.

\begin{table}
\caption{He irradiated CoFeB-samples  grown by Singulus and irradiated by Spin-Ion Technologies. The last column (\emph{Irr.}) indicates the irradiation dose.}
\label{table:implanted samples}
\small
\begin{tabular}{c|c|c|c|c}
\hline
\hline
\textbf{sample}& \textbf{FM layer} & \textbf{bottom layer} & \textbf{top layer}& \textbf{Irr.}\\
\hline
& \scriptsize{(nm)} & \scriptsize{(nm)} & \scriptsize{(nm)}  & \scriptsize{(He$^+/m^2$)}\\
\hline
\hline
\scriptsize{ID0}& \multirow{5}{*}{\centering \scriptsize{Co$_{20}$Fe$_{60}$B$_{20}$ (1)}} &  \multirow{5}{*}{\centering \scriptsize{Ta(5)}} & \multirow{5}{*}{\centering \scriptsize{MgO(2)/Ta(3)}} &\scriptsize{$0$}\\
\scriptsize{ID4}& &&&\scriptsize{$4 \times 10^{18}$}\\
\scriptsize{ID8}& &&&\scriptsize{$8 \times 10^{18}$}\\
\scriptsize{ID12}&&&& \scriptsize{$12 \times 10^{18}$}\\
\scriptsize{ID16}&&&& \scriptsize{$16 \times 10^{18}$}\\
\hline
\hline
\end{tabular}
\end{table}

To extract $D$ from bubble expansion data it is necessary to know certain magnetic parameters of the sample, which enter the model relating the DW velocity with $D$. These parameters are the saturation magnetization $M_\mathrm{s}$, the effective anisotropy constant $K_\mathrm{eff}$ and the exchange stiffness $A$.

The saturation magnetization $M_\mathrm{s}$ was measured by superconducting quantum interference device magnetometer (SQUID) (samples Tab.\ref{table:CoFeB D}), SQUID-VSM (vibrating sample magnetometer) (samples listed in Tab.~\ref{table:Co D}) and VSM (samples listed in Tab.\ref{table:irr D}). 

The anisotropy $K_\mathrm{eff}$ was obtained from the hysteresis curves for samples listed in Tabs.~\ref{table:Co samples} and \ref{table:CoFeB WPt samples}. To this aim the in-plane (hard axis) saturation field ($H_\mathrm{s}=H(m=m\pm 1/2\Delta m)$) was found by looking for the field at which the magnetic moment had changed from the saturation value by 0.5 times the $\Delta m$ error in $m$. This gives an estimate for the anisotropy field $H_\mathrm{K}$ and the effective anisotropy ($K_\mathrm{eff}$) can be calculated using $K_\mathrm{eff}=(\mu_0 H_\mathrm{s} M_\mathrm{s})/2$.

For samples in Tab.~\ref{table:implanted samples} the anisotropy was calculated by measuring magneto-optical Kerr rotation loops as a function of an in-plane magnetic field, and fitting by minimization of the energy density $E=K_\mathrm{eff} \sin^2(\theta)+K_\mathrm{eff}^2 \sin^4(\theta)-HM_\mathrm{s} \cos(\theta-\phi)$ with $\theta$ the angle between the applied field $H$ and $M_\mathrm{s}$ and $\phi$ the angle between $H$ and the easy axis \cite{HER-15}. The contribution of shape anisotropy was considered by estimating $K_\mathrm{demag}=  \frac{1}{2} \mu_0 M_\mathrm{s}^2$.
The exchange stiffness was calculated only for the Co-based samples (as shown in the \ref{sec:app}), while for the other samples literature values were used. The influence of systematic errors in the determination of these parameters on the evaluation of $D$ is discussed in detail in \ref{sec:app}.

\section{Experimental}

\subsection{General description of the asymmetric bubble expansion method}

Several experimental techniques exist to extract $D$ in FM/HM structures from the DW energy (DW methods). Among them we distinguish static (based, for example, on the direct observation of domains and DWs) and dynamic methods (based on the determination of magnetic field and electric current driven DW velocity). There are several possibilities, such as measuring domain wall velocity or energy as a function of an in-plane magnetic field, measuring domain wall spacing in stripe domain phases, or measuring the domain wall internal structure. All employ magnetic domain imaging techniques, ranging from optical to scanning force or electron microscopy (see, e.g., \cite{JE-13,CHE-13a,MEC-09a,PIZ-14,LEG-18}).

\begin{figure}[t]    
 \centering
    \includegraphics[width=\linewidth]{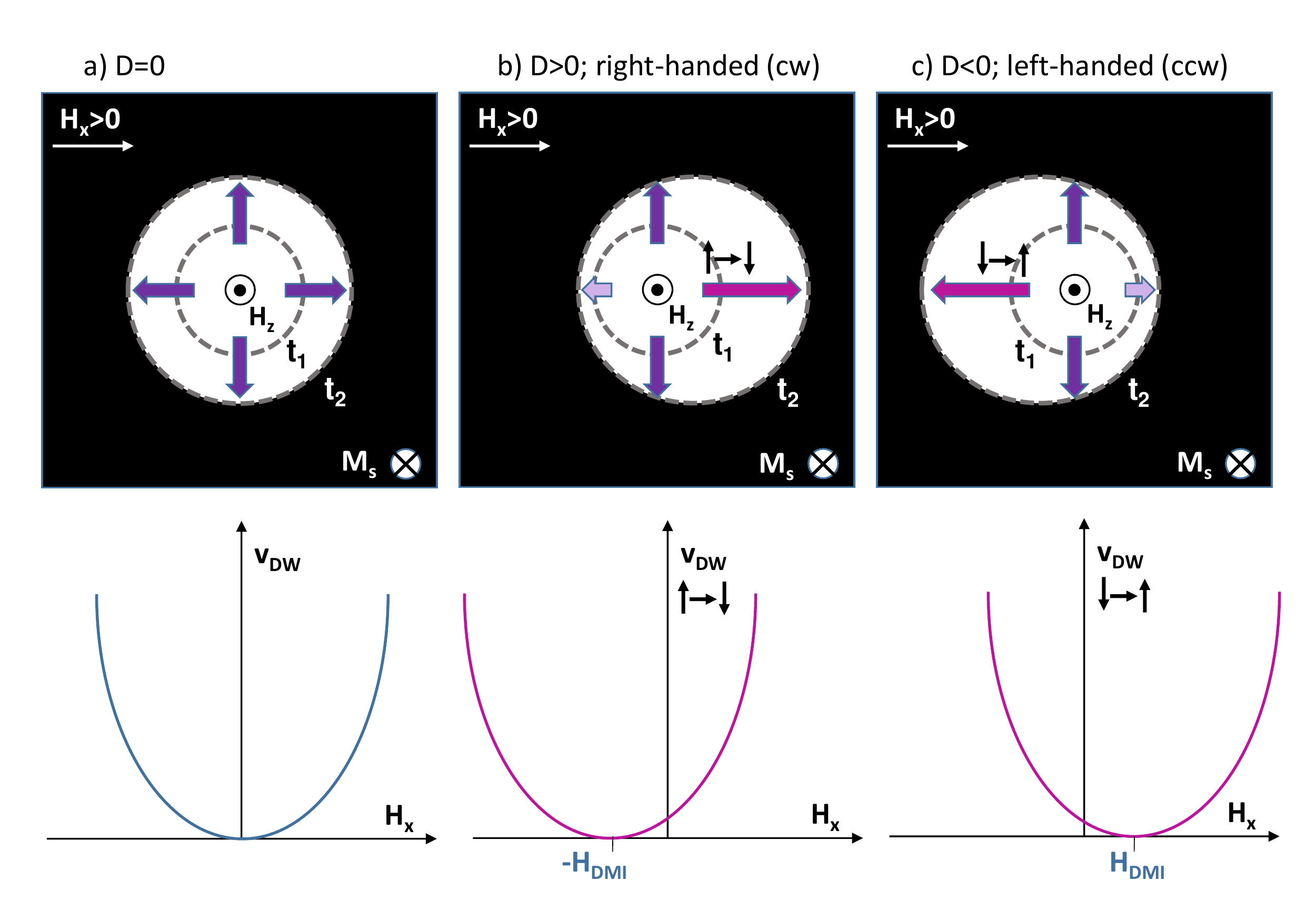}
    \caption{Conceptual illustration of the asymmetric bubble expansion for different $D$ values.  The three cases shown are (a) $D = 0$, (b) $D > 0$, favoring right-handed ($\uparrow \rightarrow \downarrow$) DWs and (c) $D < 0$, favoring left-handed ($\downarrow \rightarrow \uparrow$) DWs. Initially the film is uniformly magnetized in direction $M_s$, then a bubble DW is nucleated by a short field pulse $H_z$ (opposite to $M_s$), indicated by the grey dashed circle at time $t_1$. At a time $t_2>t_1$ the bubble has expanded to the outer grey circle. The violet/purple arrows indicate the direction and velocity of the DW motion. The lower panel shows the associated velocity curves for the DW expanding in direction of the small applied field $H_x$.
    }
    \label{fig:meas_scheme}
\end{figure}

We focus on a technique that is widely employed in the field and is available in many laboratories for characterization of magnetic thin films: the asymmetric bubble domain expansion method \cite{JE-13,KAB-10,THI-12} in the creep regime. Field-driven DW dynamics are measured mostly in continuous films or wires. In continuous films the magnetization reversal proceeds by nucleation and growth of magnetic bubble domains. The magnetization in the continuous film is initially saturated by applying a negative perpendicular field $H_z$. A bubble DW is then nucleated by applying through a coil a $H_z$ pulse in the positive direction. The bubble DW is expanded under simultaneous application of a continuous $H_x$ (from an electromagnet) and a number of positive pulses $H_z$ (from a coil). The initial and final positions of the DW are imaged by magneto-optical Kerr effect (MOKE) microscopy in polar configuration. Typical bubble growths are in the range of at least a few tens of micrometers. The velocity of the DW is measured along the direction of the applied $H_x$ (in the following we will extend the analysis to the velocity measurement along arbitrary directions) and is calculated as the ratio between the DW displacement and the total time during which $H_z$ is applied. Finally, velocities are measured for both  $\downarrow \uparrow$ and $\uparrow \downarrow$ DWs (i.e. $-x$ and $+x$ sides of the DW of a bubble) under different strength of $H_x$, whilst keeping $H_z$ constant. In this way, velocity $v_\mathrm{DW}$ versus $H_x$ curves are constructed for both DWs (see illustration of the measurement principle in Fig.~\ref{fig:meas_scheme}). The DMI field is defined as
\begin{equation}
\label{eq:HDMI}
H_{\mathrm{DMI}}=- \frac{D} {\mu_0 M_\mathrm{s} \lambda},
\end{equation}
where $M_\mathrm{s}$ is the saturation magnetization. The Bloch DW width $\lambda$ is given by 
\begin{equation}
\label{eq:lambdaBloch}
\lambda=\sqrt{A/K_{\mathrm{eff}}},
\end{equation}
where $K_{\mathrm{e}ff}$ is the effective anisotropy constant, related to the
perpendicular magnetic anisotropy constant $K$ by $K_{\mathrm{eff}}= K -\mu_0 M_s^2 /2$. The corresponding anisotropy field is 
\begin{equation}
\label{eq:anisfield}
H_\mathrm{K}=\frac{2 K_{\mathrm{eff}}}{\mu_0 M_\mathrm{s}}.
\end{equation}

The DMI field, determined through fitting with one of the creep models discussed in the following, is not necessarily found at the velocity minimum. The modeling is of substantial importance for the measurement accuracy and is discussed in section \ref{sec:Deval}.

\subsection{Magneto-optical measurements of the bubble growth asymmetry} \label{sec:exp-mopt}

Any typical wide-field magneto-optical imaging system can be used to perform the asymmetric bubble expansion measurement, provided that simultaneously both in-plane (IP) and out-of-plane (OOP) fields can be applied. INRIM uses an in-plane electromagnet with flat pole pieces able to reach $\mu_0 H_x^\mathrm{pk} \approx 150$ mT and a coil able to reach $\mu_0 H_z^\mathrm{pk} \approx 100$ mT OOP, while the University of Leeds uses an in-plane electromagnet with split pole pieces able to reach $\mu_0 H_x^\mathrm{pk} \approx 300$ mT and a coil able to reach $\mu_0 H_z^\mathrm{pk} \approx 30$ mT OOP.
A sample positioning system allows control of the sample tilt, which helps avoiding small deviations from sample planarity. Without tilt correction, a spurious OOP field component can be present when an IP field is applied.

As a first step, a position on the sample is found where a bubble can be nucleated in a repeatable way and neighboring bubbles in the field of view are as few as possible. 
The acquisition, instead of using the background subtraction commonly used in magnetooptics, is made by this sequence of steps: sample saturation, nucleation of the starting bubble, acquisition of the ``bubble image'' $I_\mathrm{B}$, bubble expansion using a square pulsed field $H_z^0$ (or sequence of pulsed fields), image acquisition $I(H_z^0)$. In this way the image $I(H_z^0)-I_\mathrm{B}$ appears to have a central hole corresponding to the starting bubble, helping to establish the value of the displacement of the domain wall. 

To ensure the sample planarity and the absence of spurious OOP fields, prior to the investigation a bubble is nucleated and expanded under both positive and negative IP field $H_x^0$. If the two bubble images are not symmetric, the sample tilt is corrected until symmetry is achieved. This ensures that the velocity curves described in the following are symmetric.

\begin{figure}    
 \centering
    \includegraphics[width=0.8\linewidth]{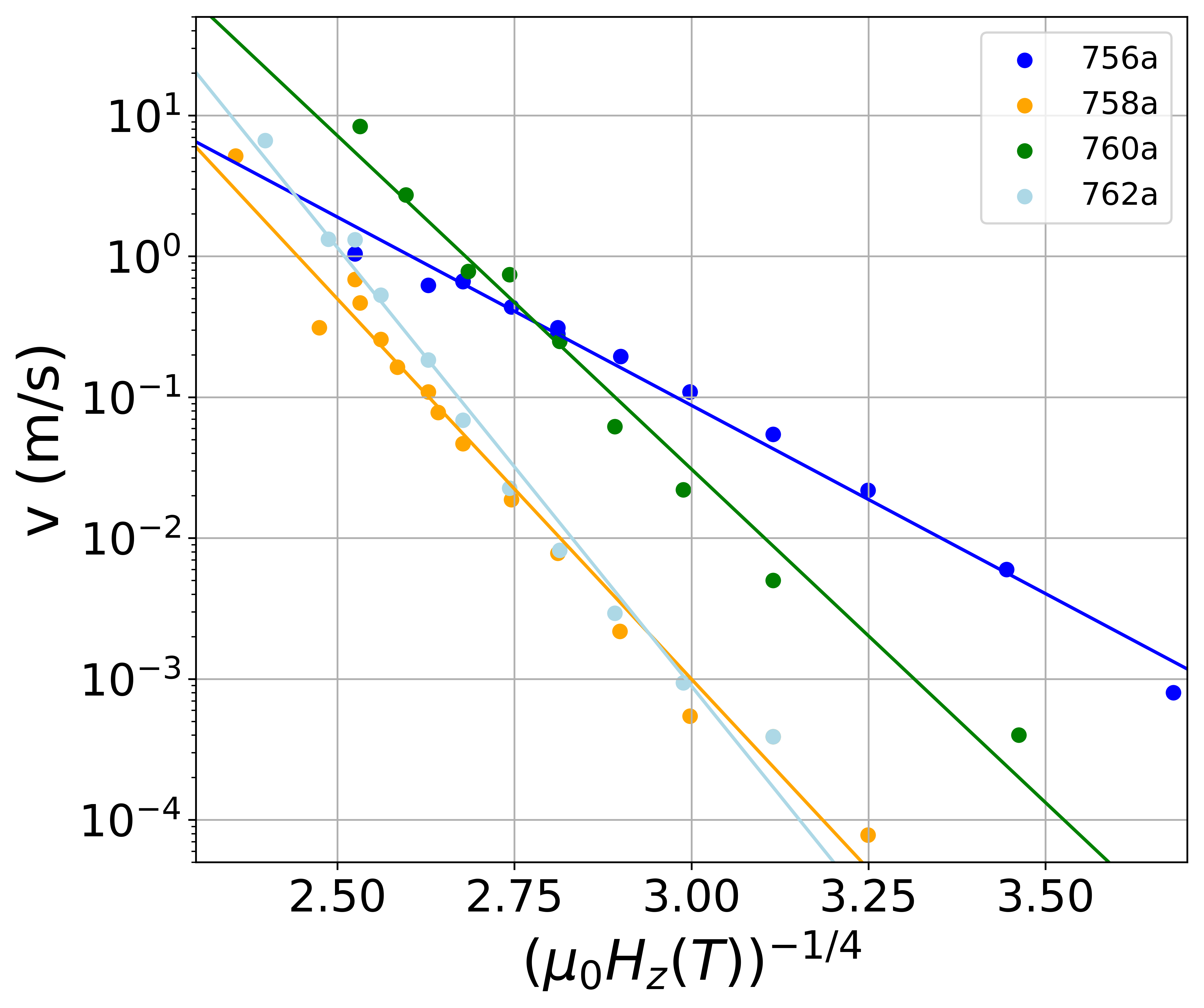}
    \caption{Creep law behavior Eq.\ref{eq:vbase} (lines) and acquired velocities (dots) for samples \textit{756a} (blue), \textit{758a} (orange), \textit{760a} (green), \textit{762a} (light blue).
    }
    \label{fig:creepbehaviour}
\end{figure}

A first set of images is acquired at zero in-plane field, under different values of the $H_z$ pulses to construct the dependence of the velocity on the out of plane field $v(H_z)$, Fig.~\ref{fig:creepbehaviour}. The base pulse duration is $\Delta t=100~\upmu$s, and multiple pulses can be used to expand the bubbles as necessary.
In the creep approximation, the power law, discussed later as Eq.~\ref{eq:vbase}, holds, and a fit allows the determination of the parameters $v_0$ and $\alpha_0$.

Having decided on a $H_z$ value well inside the creep region, both its value and its direction are kept fixed throughout all further measurement steps. Bubble images are acquired under a sequence of in-plane fields $0, \pm H_x^1, \pm H_x^2,..., \pm H_x^n$, with $H_x^1<H_x^2<...<H_x^n$ . At every new value of $H_x$ the system is saturated and the starting bubble nucleated again. Depending on the $H_x^i$ value, a different number of OOP pulse fields $H_z$ can be required to reach a high enough expansion of the bubble (under higher IP fields $H_x$ the bubble expands much faster).

\section{Evaluation of the DMI value}\label{sec:Deval}

\begin{figure}[t]    
 \centering
    \includegraphics[width=\linewidth]{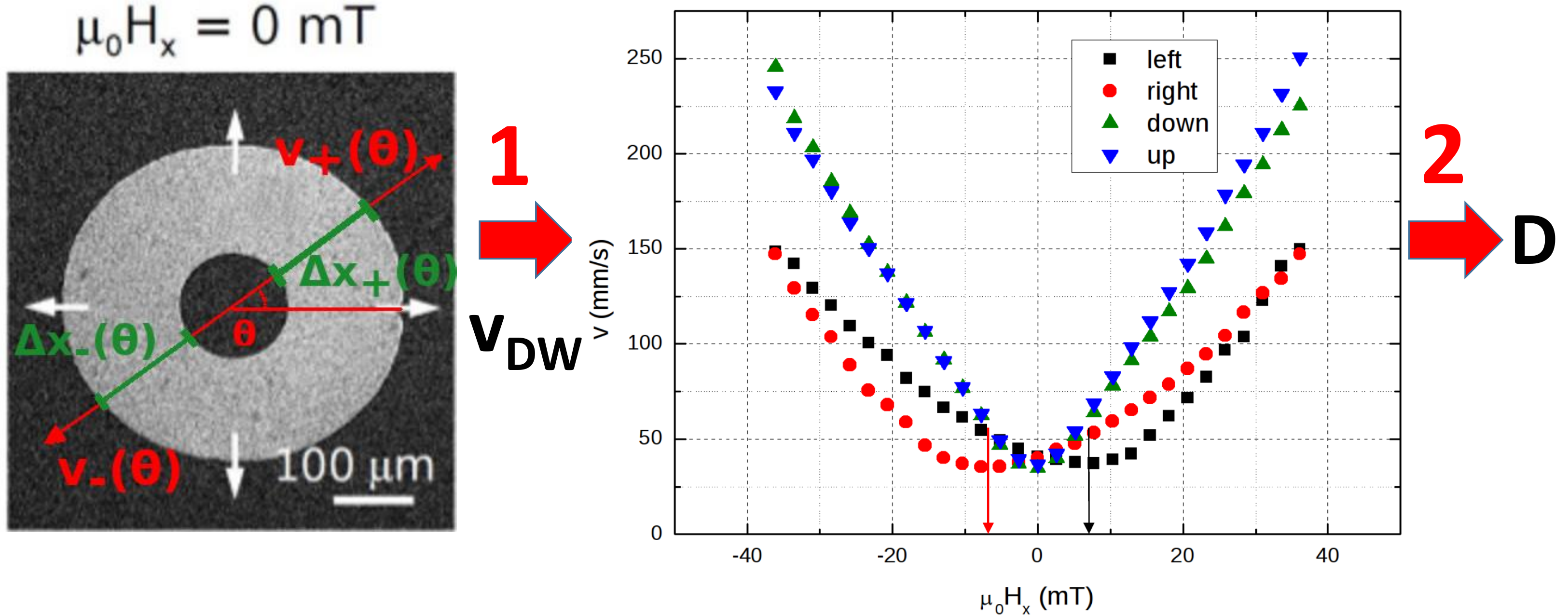}
    \caption{Representation of the two steps for evaluating $D$ (test sample: Ta(5)/Co$_{20}$Fe$_{60}$B$_{20}$(1)/MgO(2)/Ta(3) thin film grown by Singulus and irradiated by Spin-Ion Technologies): Step 1: Evaluation of the bubble domain wall velocities (right) along the in-plane field direction $\theta=0$ (black, red) and perpendicular to it $\theta=\pi/2$ (green, blue) from the MOKE image (left); Step 2: evaluation of the DMI field and $D$ from the velocity versus IP field curves.}
    \label{fig:evaluation}
\end{figure}

The evaluation of the DMI value $D$ from the measured data, i.e., the sequence of magneto-optical images of the bubble domain expanding asymmetrically, can be divided into two steps: first, extraction of the DW velocities as a function of applied in-plane field $H_x$, and second, evaluation of the field where the DW velocity is minimum, which will deliver the DMI field $H_\mathrm{DMI}$ (see Fig.~\ref{fig:evaluation}), at least within the simplest picture. Regarding the second step, we focus on the creep regime and therefore creep models are employed for the evaluation of the velocity minimum. Several models exist in the literature, but a classification, describing to which kind of heterostructures or materials a model can be applied with success, is missing. Regarding the first step, in cases of rough bubble domain walls (see Fig.~\ref{fig:PtMLCoFeBDomains}) the evaluation of the DW velocity might be tricky or even not possible. Furthermore, sometimes the bubbles do not expand fastest exactly in direction of the applied in-plane field, so an integrated view of the DW velocities along all directions around the starting bubble can give more information to obtain a valid DMI measure.  It is also noteworthy that internal domain wall dynamics in some cases have been shown to influence the steady state magnetization profile leading to growth directions that deviate significantly from the in-plane field axis \cite{JAB-01}.

\subsection{Extraction of the DW velocity}

\subsubsection{Image Processing}
The bubble images acquired as described in section \ref{sec:exp-mopt} are processed in contrast, brightness, and blur, to ease the automatic localization of the DW position.
Starting from the bubble central position, selecting a direction at an angle $\theta$ with the $H_x$ direction, we find the distances $\Delta x_-(\theta)$, $\Delta x_+(\theta)$ traveled by the DWs in the $-x$ and $+x$ directions respectively (as indicated in Fig.~\ref{fig:evaluation}, left). This is accomplished by a software procedure that checks for major jumps in the image contrast. Neighboring bubbles can generate errors in the processing, but the spurious data are easily identified and manually discarded. The $\Delta x_-(\theta)$, $\Delta x_+(\theta)$ values are re-scaled by the pixel size, depending on the camera and the objective used. Finally, the velocities $v_-(\theta)$, $v_+(\theta)$ (among them the velocities along $H_x$: $v_{\downarrow \uparrow}=v_-(0)$ and $v_{\uparrow \downarrow}=v_+(0)$) are calculated dividing $\Delta x_-(\theta)$, $\Delta x_+(\theta)$ by the total pulse duration (the single pulse has fixed duration $\Delta t = 100~\upmu$s, but different numbers of pulses are used, depending on $H_z$ intensity).

\subsubsection{Alternative image processing: The MOKAS software}

\begin{figure}[t]
 \centering
    \includegraphics[width=\linewidth]{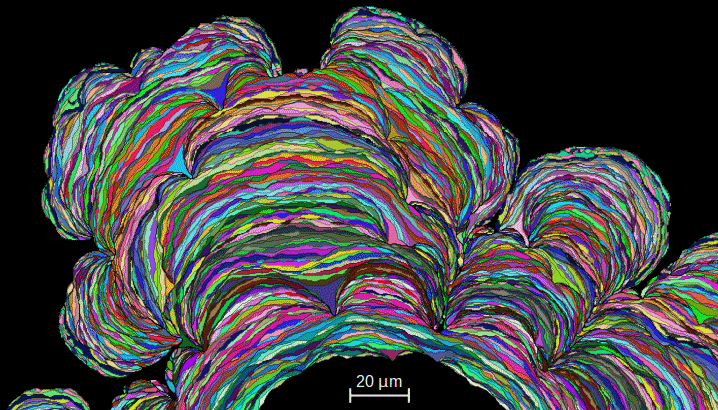}
    \caption{Illustrative example of a bubble expansion in presence of many pinning centers, as calculated by the MOKAS software. The initial bubble, nucleated at the beginning of the experiment, is partially visible in the lower part of the image in the center (in black). The expansion is represented by the bubbles of various colors, each color representing a the evolution of the bubble at a new time step. Sample ID16, Tab.\ref{table:implanted samples}.}
    \label{fig:mokasexample}
\end{figure}

A more sophisticated (but less straightforward) way to extract the DW velocity, especially suitable in case of low contrast and rough bubbles, is to use an ad hoc analysis software applied to a sequence of MOKE images (in form of a video of the bubble expansion) based on the detection of the image contrast (i.e., the gray level) change. This software is named \textsc{MOKAS}, is freely available on GitHub.com (\doi{10.5281/zenodo.5714377}, \href{https://github.com/gdurin/mokas}{MOKAS software}), and uses parallel computing to estimate the time frame at which a single pixel of the images changes the gray level as a consequence of the motion of the DW wall. In other words, the bubble shape is calculated at each time frame and thus the velocity of the entire contour can be extracted, see example in Fig.~\ref{fig:mokasexample}. 

\subsection{The sign of $D$}

The bubble expansion method is able to determine the sign of the $D$ value and therefore the DW chirality (or in certain cases even the DW type, Bloch or Néel). The chirality can be already obtained from the direction of the major bubble expansion without the need to extract the DW velocity. Care has to be taken only in considering correctly the signs of the applied fields, since inverting applied fields means to invert the direction of expansion. We observe that in the Pt-based samples of Tabs.~\ref{table:Co samples} and~\ref{table:CoFeB WPt samples} we find $v_{\uparrow \downarrow}<v_{\downarrow\uparrow}$ for $H_x>0$ while the opposite is true for $H_x<0$.
This suggests a negative value of the DMI value, because it corresponds to counterclockwise domain walls (left hand chirality).
The opposite sign ($D>0$) is found instead in the W-based samples of Tab.~\ref{table:CoFeB WPt samples} and in the irradiated samples in Tab.~\ref{table:implanted samples}. This is in agreement with results obtained by other methods \cite{VAN-15, SOU-16, MA-18}.

\subsection{Modeling the DW velocity}
The simplest model of asymmetric bubble expansion is based on fitting a parabola to the $v(H_x)$ curve and to assume that $H_\mathrm{DMI}$ is the value of the minimum, where $H_x$ exactly cancels $H_\mathrm{DMI}$.
The next level of sophistication is to use the basic creep model \cite{JE-13,HRA-14} in which the effect of $H_x$ on the wall energy is accounted for. This has the merit of explaining one of the main features of the domain wall velocity in presence of DMI, i.e., the fact that the velocity minimum does not occur at $H_x = 0$ when there is finite $D$. It also reliably gives the sign of $D$. However, the basic creep model is not able to model a large number of materials and bubbles which show important deviations from a rigid shift of the creep curves along the $H_x$ axis by $-H_\mathrm{DMI}$.
In particular, $v(H_x)$ often has a pronounced asymmetry about its minimum \cite{LAV-15,JUE-16a}. Furthermore, in some cases the velocity curves are extremely flat (see e.g. Fig.\ref{fig:JeFitLeeds}). Sometimes, especially for irradiated samples, a clear change of slope occurs at higher fields \cite{LAV-15,JUE-16a,KIM-16d,VAN-15}. 

To address these problems several extensions were proposed, which in some cases obtain good results in modeling distorted velocity curves, as the micromagnetic model with field-dependent depinning field \cite{SHA-19}, or the chiral damping model \cite{JUE-16a}, where a dissipative spin-orbit torque on the DW dynamics explains the asymmetry in the energy density.
Here we will consider the arbitrary angle propagation model \cite{KIM-17}, the creep model extension with varying DW width \cite{KIM-16}, and the dispersive stiffness model \cite{LAU-16,PEL-17,LAU-18}. 


The issue of asymmetry in the energy density is characteristic for measurements in the creep regime, while it is absent  for DW motion in the flow regime \cite{KIM-16d,VAN-15}. Nevertheless, here we focus on the creep analysis, since this regime has been widely investigated in the literature, and has the potential to become a standard technique due to the simpler experimental procedure.

\subsection{Basic creep model}

The DMI energy in perpendicular materials stabilizes Néel DWs with a fixed chirality, with the rotation sense given by the sign of the DMI value \cite{THI-12}.  This energy can be considered as an additional field $H_\mathrm{DMI}$ acting on the DW, and its value can be obtained by examining the domain asymmetric motion under an additional in-plane field $H_x$ \cite{JE-13,HRA-14}. As stated above, the simplest assumption is that the value of in-plane field $H_x^0$, where the velocity minimum occurs, balances the DMI field, and so $H_x^0=-H_\mathrm{DMI}$. This then yields the value of $D$, Eq.\ref{eq:HDMI} \cite{THI-12}.

The DW motion is studied in the creep regime \cite{MET-07}, as a competition between DW elasticity and material disorder, where the velocity is given by:
\begin{eqnarray}
\label{eq:vbase}
v & = &v_0 \exp \left( - \alpha H_{z}^{-1/4}  \right) \nonumber \\
& = &v_0 \exp \left( - \alpha_0 \epsilon^{1/4} H_{z}^{-1/4}  \right) \nonumber \\
& =& v_0 \exp \left( -\alpha_0 [\sigma(H_{x})/\sigma(0) ]^{1/4} H_z^{-1/4} \right).
\end{eqnarray}
Here $H_z$ is the OOP field driving the bubble growth and $v_0$ its characteristic speed; the scaling parameter is $\alpha \propto \epsilon^{1/4}$ where $\epsilon$ describes the potential associated with the bending deformation of the interface and is here identified with the DW energy density. We can therefore write $\alpha=\alpha_0 \sigma(H_{x})/\sigma(0)$ with $\alpha_0$ a scaling factor independent of $H_x$. 

The power law described by Eq.~\ref{eq:vbase} is valid only in the so-called creep regime; once the $H_z$ field reaches a threshold value, the system transitions into the flow regime \cite{MET-07,SHA-19}.
By measuring the $v$ dependence on $H_z$, see Fig.~\ref{fig:creepbehaviour}, we are able to determine $v_0$, $\alpha_0$ in Eq.~\ref{eq:vbase}.
In the basic creep model the DW energy density also depends on the DMI field, and is written as:
\begin{equation}
\label{eq:sigmaDWdef}
\sigma(H_x,\psi)=\sigma_0 + 2 K_\mathrm{D} \lambda \cos^2 \psi  - \pi \mu_0 M_s \lambda (H_x+H_\mathrm{DMI})\cos \psi,
\end{equation}
where the terms on the right are the individual energy density contributions: Bloch DW, DW anisotropy, Zeeman, and DMI. The basic model is valid only for the points of the wall where the DW normal $\hat{n}$ is parallel to the field $H_x$. The angle $\psi$ is defined as the angle between the field $H_x$ (or the DW normal) and the magnetization, Fig.~\ref{fig:arbangle} (left).
$H_\mathrm {DW}=4K_\mathrm{D}/\pi\mu_0 M_\mathrm{s}$ is the DW anisotropy field, and  $K_\mathrm{D}=\ln(2)\ t \mu_0 M_\mathrm{s}^2 / 2 \pi \lambda$ is the DW anisotropy energy density, with $t$ the magnetic film thickness;
the Bloch DW energy density is given by $\sigma_\mathrm{DW}=4 \sqrt{A K_\mathrm{eff}}$. 

Since $\sigma$ depends upon the magnetization direction $\psi$, we can obtain the value $\sigma(H_x)$ by finding the equilibrium magnetization angle $\psi_\mathrm{eq}$:
\begin{equation}
\cos \psi_\mathrm{eq} =\pi \frac{M_\mathrm{s} (H_x+H_\mathrm{DMI})}{4K_\mathrm{D}},
\end{equation}
so that Eq.~\ref{eq:sigmaDWdef} admits the two possible solutions:
\begin{eqnarray}
\label{eq:sigmaDW1}
\sigma &=&\sigma_0+2K_\mathrm{D} \lambda - \pi \lambda \mu_0 M_\mathrm{s} |H_x+H_\mathrm{DMI}|,~\mathrm{or}
\\
\label{eq:sigmaDW2}
\sigma &=& \sigma_0-\frac{\pi^2 \lambda \mu_0^2 M_\mathrm{s}^2}{8K_\mathrm{D}}(H_x+H_\mathrm{DMI})^2.
\end{eqnarray}
Eq.~\ref{eq:sigmaDW1} is valid for a pure Néel DW ($\psi_\mathrm{eq}=0,\pi$, when $H_x+H_\mathrm{DMI}>H_\mathrm{DW}$),
while Eq.~\ref{eq:sigmaDW2} is valid for a hybrid Bloch-Néel DW ($0<\psi_\mathrm{eq}<\pi$, when $H_x+H_\mathrm{DMI}<H_{DW}$). According to this model, with Eqs.~\ref{eq:sigmaDW1},\ref{eq:sigmaDW2} inserted into the velocity formula Eq.\ref{eq:vbase}, the velocity curves show inversion symmetry with respect to $H_x^0=-H_\mathrm{DMI}$, and the DMI value $D$ can be extracted from the location of the symmetry axis.

\subsection{Arbitrary angle propagation}

In Ref.~\cite{KIM-17}, the basic creep model is extended to include the propagation of DWs with the DW normal at an arbitrary angle with respect to $H_x$. So instead of measuring the DW velocity along the $H_x$ direction, we will determine its velocity over a range of angles.

\begin{figure}[t]    
 \centering
    \includegraphics[scale=0.5]{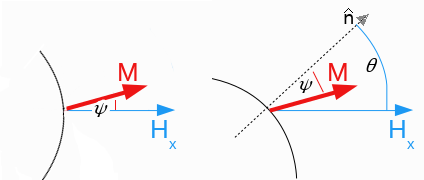}
    \caption{Definition of angles used in Eq.~\ref{eq:sigmaDWdef} (left) and Eq.~\ref{eq:sigma_arbangle} (right) for an arbitrary DW segment (curved black line)}
    \label{fig:arbangle}
\end{figure}

The energy density for arbitrary DW orientation is given by:
\begin{eqnarray}
\label{eq:sigma_arbangle}
\sigma( H_x,\psi ) &=&\sigma_0+2 K_D \lambda \cos^2\psi - \pi \mu_0 M_\mathrm{s} \lambda \times \nonumber\\
& &  \left[ ( H_x \cos \theta  + H_\mathrm{DMI}) \cos\psi + \right. \nonumber \\ & & \left. H_x \sin \theta  \sin \psi \right],
\end{eqnarray}
with the angles $\theta$ and $\psi$ shown in Fig.~\ref{fig:arbangle} (right).

The minimization $\partial \sigma / \partial \psi =0 $ gives the equilibrium position $\psi_\mathrm{eq}$, while the maximization $\partial \sigma / \partial H_x =0 $ gives the maximum position $H_x=H_0$ of $\sigma$ (so the $v$ minimum):
\begin{equation}
\label{eq:H0_arbangle}
H_0= \left( \pm H_\mathrm{DW} \sin \theta - H_\mathrm{DMI}  \right) \cos \theta.
\end{equation}
The exploration of the velocity minimum as a function of the angle $\theta$ then allows us to obtain the $H_\mathrm{DMI}$ and $H_\mathrm{DW}$ values by fitting the data to Eq.~\ref{eq:H0_arbangle}.

\subsection{Varying DW width}

In Ref.~\cite{KIM-16}, a modified creep model is developed, following the observation that the DMI must introduce a variation of the domain wall width $\lambda$, up to now given by Eq.~\ref{eq:lambdaBloch}, which is in fact correct just for Bloch DWs, whereas the DMI modifies their structure. By introducing this variation, the DW energy density $\sigma_{DW}(H_x)$ itself contains an asymmetric contribution.
By making $\lambda$ explicit we have:
\begin{eqnarray}
\sigma( H_x, \psi , \lambda) & = & 2(A/\lambda+K \lambda) + 2 K_\mathrm{D} \lambda \cos^2 \psi - \nonumber \\
& & \pi \mu_0 M_\mathrm{s} \lambda ( H_x+H_\mathrm{DMI} ) \cos \psi.
\label{eq:sigmaVaryingWidth}
\end{eqnarray}
By minimization, both $\psi_\mathrm{eq}$ and $\lambda_\mathrm{eq}$ fall into three regimes that correspond to each of three different DW structures. With increasing $H_x$ the system passes in sequence through the stages NW$^-$ ($\hat m || H_x$), BW-NW, NW$^+$ ($\hat m || -H_x$ ). In the NW$^+$ and NW$^-$ structures, the magnetization inside the DW is saturated along the $+x$ and $-x$ axes respectively (Néel DWs, $\psi_\mathrm{eq}=0,\pi$), Eq.~\ref{eq:sigmaeqvaryingwidth1}; the BW-NW structure instead corresponds to the transition between Bloch and Néel-type DWs, Eq.~\ref{eq:sigmaeqvaryingwidth2}.

Once the equilibrium values of $\psi_\mathrm{eq}$ and $\lambda_\mathrm{eq}$ are known, substituting them in the energy density definition Eq.~\ref{eq:sigmaVaryingWidth} we obtain:
\begin{equation}
    \label{eq:sigmaeqvaryingwidth1}
    \sigma_\mathrm{eq}(H_x,\psi_\mathrm{eq},\lambda_\mathrm{eq})=4\sqrt{A(K-\pi \mu_0 M_\mathrm{s} H_x/2)}+(\ln 2/\pi)\ t\   \mu_0 M_\mathrm{s}^2 \mp \pi D,
\end{equation}
and
\begin{equation}
    \label{eq:sigmaeqvaryingwidth2}
    \sigma_\mathrm{eq}(H_x,\psi_\mathrm{eq},\lambda_\mathrm{eq})=2\left(\frac{A}{\lambda_\mathrm{eq}^\mathrm{BN}}+K \lambda_\mathrm{eq}^\mathrm{BN} \right)-\frac{\pi^3}{4 \ln 2} \frac{\left( \lambda_\mathrm{eq}^\mathrm{BN} \mu_0 M_\mathrm{s} H_x+D \right)^2}{d \mu_0 M_\mathrm{s}^2},
\end{equation}
with Eq.~\ref{eq:sigmaeqvaryingwidth1} valid for NW$^\pm$ domain walls, Eq.~\ref{eq:sigmaeqvaryingwidth2} valid for BW-NW domain walls, and $\lambda_\mathrm{eq}^\mathrm{BN}$ being the equilibrium DW width in the BW-NW regime. $H_x$ is asymmetric with respect to both $\psi_\mathrm{eq}$ and $\lambda_\mathrm{eq}$, which consequently leads to an asymmetry in the equilibrium DW energy density $\sigma^\mathrm{eq}$. This in turn generates an asymmetry in the velocity curves of $v_{\uparrow \downarrow},v_{\downarrow\uparrow}$. The origin of this asymmetry is mainly the dependence on $\lambda$ of $H_\mathrm{DMI} \approx D/\lambda$, giving a nonlinear contribution to the effective field $H_x+H_\mathrm{DMI}$.

\subsection{Dispersive stiffness}

The relationship between $v$ and $\sigma$ at the base of Eq.~\ref{eq:vbase} is discussed in \cite{PEL-17,LAU-18,LAU-16}.
Behind the identification $\alpha \propto \epsilon^{1/4}=\sigma^{1/4}$ made by the standard creep model lies the assumption that $\sigma$ does not depend on the DW orientation. However, it was shown in Eq.~\ref{eq:sigma_arbangle} that $\sigma$ is a function of $\theta$ in the most general case. The dispersive stiffness model sets $\epsilon$ equal to the surface stiffness $\tilde \sigma$, depending on the energy of the local in-plane orientation $\theta$, and also on the energies of orientations in close proximity to $\theta$: $\tilde{\sigma}(\theta)=\sigma(\theta)+\sigma''(\theta)$, with $\sigma''(\theta)=0$ if the DW energy is isotropic. This line of argument yields
\begin{equation}
\label{eq:stiffvelocity}
v=v_0 \exp \left(-\alpha_0 \left( \tilde{\sigma}(H_x)/\tilde{\sigma}(0)  \right)^{1/4}  H_z^{-1/4} \right).
\end{equation}
Considering the DW energy density for an arbitrary orientation Eq.\ref{eq:sigma_arbangle}, we can calculate the stiffness value:
\begin{equation}
\label{eq:sigmastiff}
\tilde{\sigma}(\theta)=\sigma(\theta)+\sigma_{\theta \theta} - \frac {\sigma^2_{\theta \psi}} {\sigma_{\psi \psi}} \zeta(L/2 \Lambda),
\end{equation}
with the convention $\sigma_{xy}=\partial^2 \sigma / \partial x \partial y$ and $\zeta(x) = 1-3(x-\tanh(x))/x^3$. $\Lambda$ is the exchange length along the domain wall, and the parameter $L$ describes the DW deformation length scale.

In the limit $L\rightarrow0$ (long wavelength distortion limit) the stiffness value becomes:
\begin{equation}
\tilde{\sigma}(\theta)=\sigma(\theta)+\partial^2\sigma(\theta)/\partial \theta^2.
\end{equation}
In this limit, the stiffness corresponds to the domain wall bending while maintaining a constant internal magnetization direction.  Two important consequences of this model are 1) the minimum in growth velocity does not typically coincide with $H_\mathrm{DMI}$ and 2) the velocity of the left and right sides of the bubble converge for large in-plane magnetic fields.  It is noteworthy that the comprehensive micromagnetic modeling of Shahbazi et al \cite{SHA-19} arrives at a similar result for systems with built in magnetic disorder.



\section{Results and discussion}

\subsection{Domain structures and velocity curve shape}

\begin{figure}[t]    
 \centering
    \includegraphics[width=\linewidth]{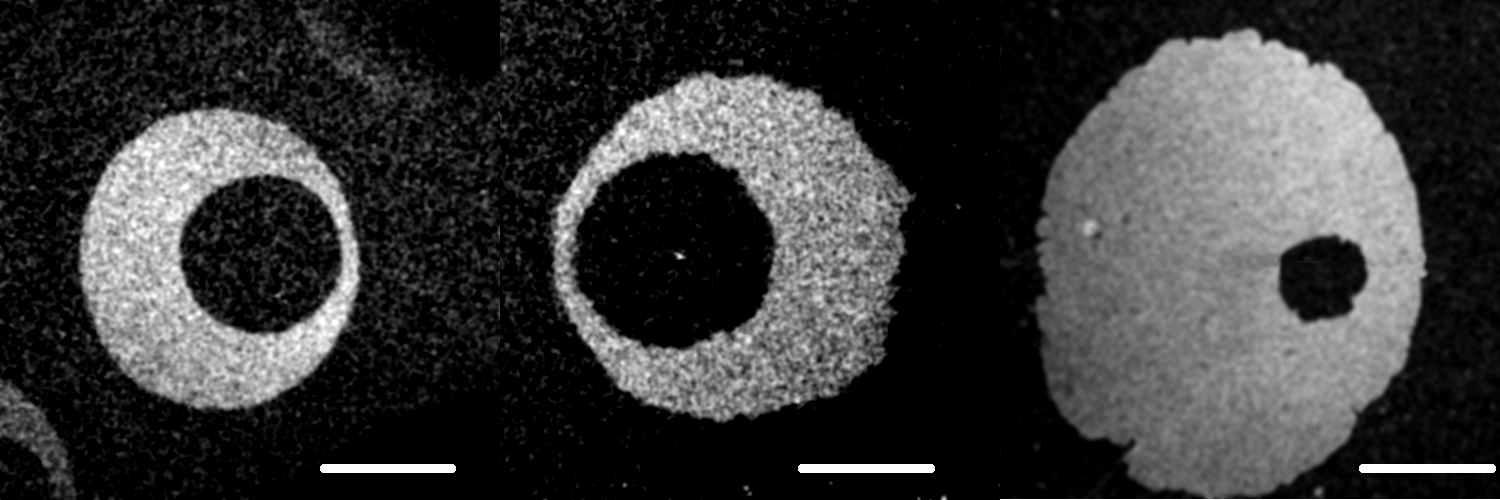}
    \caption{Regular bubbles expanding under different $H_x$ values, in W/FeCoB 
    sample 756a (left),
    Pt/Co sample a5 (center) 
    and irradiated sample ID16 (right). Scale bar is $100~\upmu$m long.
    \label{fig:regularBubbles}}
\end{figure}

\begin{figure}[t]    
 \centering
    \includegraphics[width=0.7\linewidth]{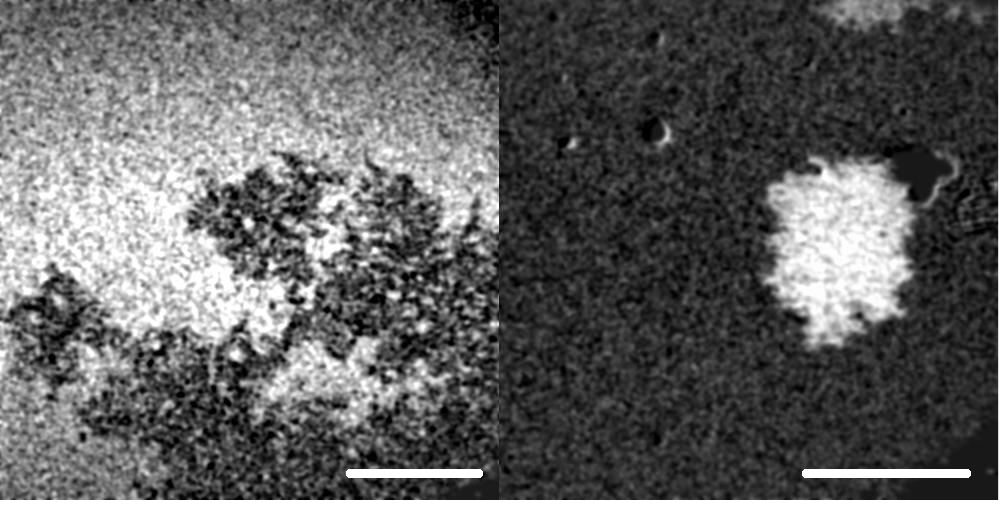}
    \caption{Irregular domain structures, where the bubble expansion analysis can be hard or impossible. Pt/CoFeB annealed multilayered samples 759a  (left), and 763a (right). Scale bar is $100~\upmu$m long.}
    \label{fig:PtMLCoFeBDomains}
\end{figure}

\begin{figure*}[t]    
 \centering
    \includegraphics[width=0.95\linewidth]{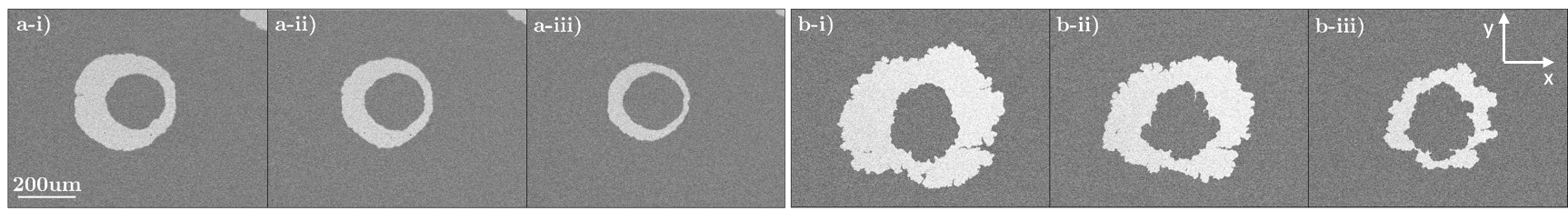}
    \caption{a(i-iii): W/CoFeB sample 756a, annealed. Expansions i-iii used 38~Oe OOP field pulses with durations of 2~s, 1.5~s, and 1s respectively, in the presence of a 175~Oe IP field.  b(i-iii): W/CoFeB sample 756b as grown, no anneal. Expansions (i-iii) used 13~Oe OOP field pulses with durations of 2.5~s, 2~s, and 1~s respectively, all in the presence of a 100~Oe IP field. In each sample images are taken at one nucleation site, with new bubbles nucleated in the centre of each image for each expansion. While the nucleated bubble is identical in a(i-iii), its shape changes each time in b(i-iii) in a non repeatable manner.
}
    \label{fig:WCoFeB}
\end{figure*}

The different sets of samples have different qualities of bubble domains. Examples for the different bubble domain qualities are shown in Fig.~\ref{fig:regularBubbles}. The domain walls are slightly rough in the Pt/Co/Pt samples (Tab.~\ref{table:Co samples}), whereas well-defined smooth bubbles are found in both the Pt/Co/Ir and Pt/Co samples (Fig.~\ref{fig:regularBubbles} center). The domain structures in the W/CoFeB/MgO samples (Tab.~\ref{table:CoFeB WPt samples}) consist of rough bubbles Fig.~\ref{fig:WCoFeB}(b), which transform to smooth bubbles under annealing, Fig.~\ref{fig:WCoFeB}(a). 
In the multilayered samples with corresponding composition the structures are rough bubbles, with long pinning lines.
In the Pt/CoFeB/MgO samples smooth bubbles are observed. The corresponding multilayers with several repetitions show instead a transition from a maze structure at low Pt deposition power (sample 759a), Fig.~\ref{fig:PtMLCoFeBDomains}(left), to rough bubbles at high Pt deposition power (sample 763a), Fig.~\ref{fig:PtMLCoFeBDomains}(right).
The irradiated samples (Tab.~\ref{table:implanted samples}) exhibit well-defined bubbles Fig.~\ref{fig:regularBubbles} (right), where samples irradiated with a fluence lower than $12\times10^{18}$~He$^+$/m$^2$ display a visible elongation in the vertical direction under small $H_x$ fields. This has been previously observed \cite{DIE-19a}, because systems with small DMI have DWs in the Bloch configuration, the two Bloch chiralities are degenerate, and the application of $H_x$ reinforces the Bloch configuration for the $\theta=\pm\pi/2$ DWs, making them expand faster.

The DMI analysis depends upon a correct determination of the DW velocity and therefore the shape and quality of the bubble domain studied is of utmost importance. For bubbles with a rough DW the precise evaluation of the DW velocity is extremely difficult. If the domain structure is close to the transition to a maze structure, the method is not applicable. Moreover, care must be taken to avoid many bubbles generated too close to each other in order to prevent strong magnetostatic interactions among neighboring bubbles which would act as a brake on the domain wall expansion and lead to a measurement of DW velocity depending on an additional, not controlled effect.

Furthermore, certain models apply well to to certain types of DW velocity curves, as discussed below. An important point is to understand if the bubble shapes (smooth or rough, regular or irregular) and types of expansion (in direction of $H_x$, perpendicular to it or at an arbitrary angle) are sufficient to predict the shape of the velocity curve and therefore the model which has to be applied. Fig.~\ref{fig:WCoFeB} shows both W/FeCoB samples, as grown and annealed. The as-grown sample has a very rough bubble domain wall Fig.~\ref{fig:WCoFeB} (b), and exhibits a flat velocity curve Fig.~\ref{fig:JeFitWCoFeB} (left), while the annealed sample domain wall is smoother Fig.~\ref{fig:WCoFeB} (a), with a more parabolic velocity curve Fig.~\ref{fig:JeFitWCoFeB} (right).
Fig.~\ref{fig:WCoFeB}~b(i-iii) shows bubble expansions performed on the only as-grown sample having sufficient OOP anisotropy to produce visible bubbles. It is shown that the bubble nucleated before the asymmetrical expansion is not forming in a repeatable manner, with significantly different shapes and sizes despite the use of the same OOP nucleation field and site for each image. Compared to the sequence of images produced in the annealed sample, Fig.~\ref{fig:WCoFeB}~a(i-iii), it is clear that the not-annealed samples are not as well suited to the technique.

Although the rough shape sometimes leads to difficulties in the determination of the velocity minimum here a reasonable measurement error was obtained by the standard creep model (see Tab.~\ref{table:CoFeB D}, samples 756a and 756b). Also the agreement between the measurements performed at University of Leeds and INRIM and with independent measurements performed by BLS is good (see Fig.~\ref{fig:RRD}), even though the $D$ values are small. It is also worth noting that the annealing step has an important role regarding the easy axis of magnetization for the samples. Often, as grown samples have a hard magnetization axis OOP, whereas after annealing they develop an easy magnetization axis OOP, and the asymmetrical bubble expansion technique is only usable for strongly OOP samples where the IP field applied to produce the asymmetry in the bubbles is not large enough to start to magnetize the samples in-plane during the expansions. 

The annealed Pt/CoFeB samples have all much flatter velocity curves and rougher bubble domain walls than the annealed W/FeCoB sample. In presence of flat curves, especially for small $H_\mathrm{DMI}$ values, the fitting of the DMI field becomes critical, and the error on its estimate rises substantially. In fact, the measurement error is rather large for all samples (up to 40\% error by using the standard creep model). Often, rough bubbles have flatter velocity curves than smooth bubbles; however, no clear trend concerning the bubble roughness with the sputter power can be observed. With care, considering the large errors, we may state that $D$ increases with the sputter power (see Table \ref{table:CoFeB D}).

Much smoother bubble domain walls occur for the He$^+$ irradiated samples with slight irregularities in the bubble domain circumference with increasing irradiation dose. An increase in interface width was reported in \cite{HER-19} with He$^+$ irradiation yielding an increase in $D$. One has to consider that the irradiation changes $M_s$ as well as $K_\mathrm{eff}$, two parameters that enter the fit for obtaining $H_\mathrm{DMI}$ and the evaluation of $D$ from $H_\mathrm{DMI}$.

\begin{figure}[t]
 \centering
    \includegraphics[width=\columnwidth]{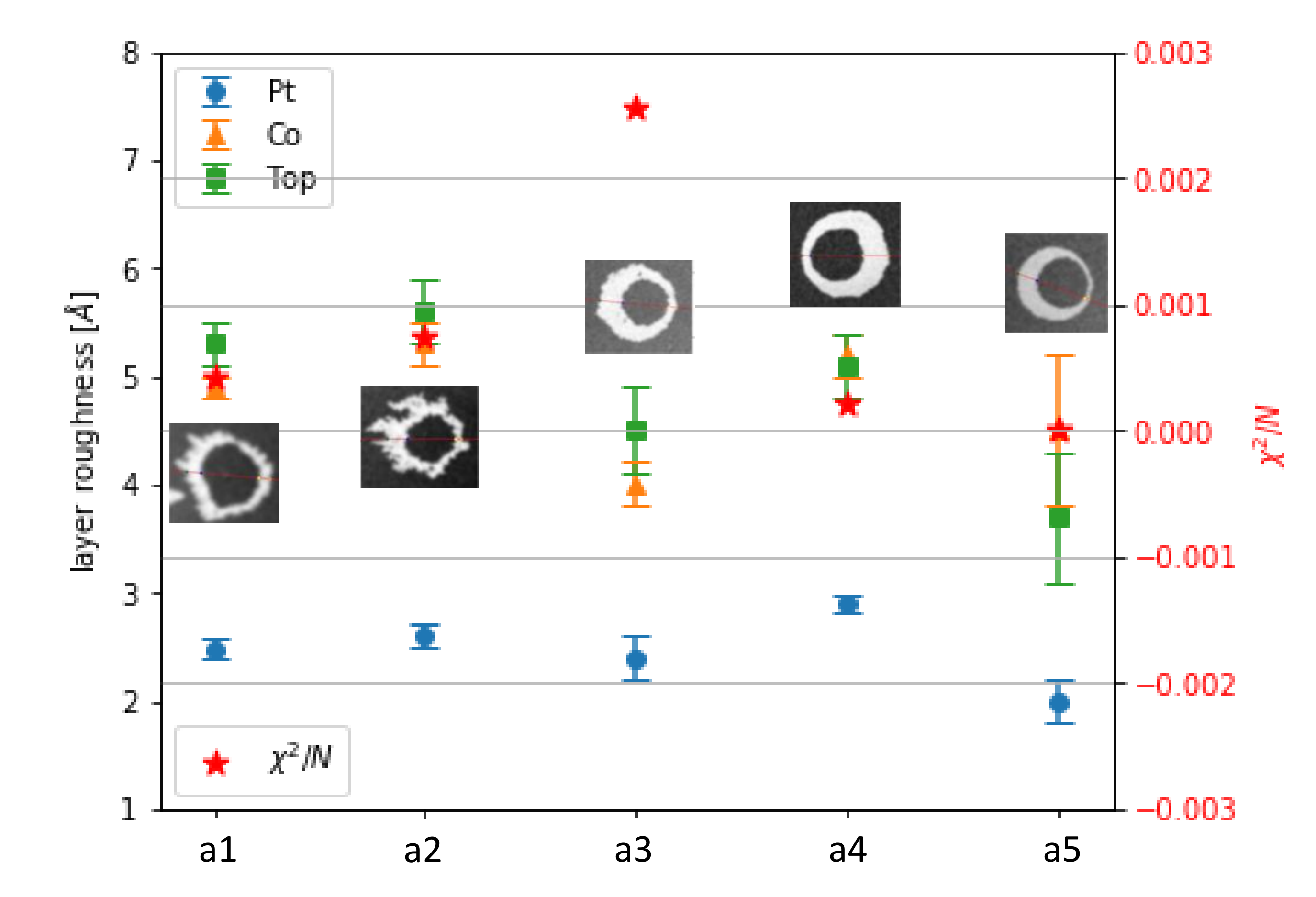}
    \caption{Co-based samples (a1-a5): Comparison of layer roughness obtained from GenX fits of low angle XRD, bubble domain wall roughness and deviation of the velocity curve shape from a parabola, given by $\chi^2/N=\frac{1}{N}\sum (v_\mathrm{meas}-v_\mathrm{fit})^2$.}
    \label{fig:roughness}
\end{figure}

The Co-based samples have medium rough bubble domain walls and their velocity curves are rather difficult to be fitted by the standard creep model. We compare bubbles and velocity curve shape with the fitted layer roughness obtained from low angle XRD (see Fig. \ref{fig:roughness}). As an estimate for the deviation of the velocity curve from a parabolic shape we use $\chi^2/N=\frac{1}{N}\sum (v_\mathrm{meas}-v_\mathrm{fit})^2$, where $N$ is the number of measured velocity points, $v_\mathrm{meas}-v_\mathrm{fit}$ is the difference between measured velocity at a given $H_x$ and fitted one. Although differences between the samples are small we find a correlation between bubble DW roughness, top layer roughness and velocity curve shape, confirming the hypothesis that rougher DWs lead to flatter velocity curves (or curves which deviate more from a parabola). Furthermore, the DW roughness seems to be higher the higher the layer roughness.

\subsection{Application of the various models and $D$ values}

The standard creep model is applied by a non-linear least squares fit with Eq.\ref{eq:vbase} on the data from both $v_{\uparrow \downarrow}$ and $v_{\downarrow \uparrow}$. Since points at different fields have been acquired using a different number of pulses (with the base pulse length being $T=100~\upmu$s) we have for the \textit{i-}th pulse the duration $t_i=n_i \cdot T$, so we set the uncertainty in the velocity of the individual points acquired to $s(v_i)=w/t_i$, where $w$ is the pixel width and $t_i$ the \textit{i}-th pulse duration.
We choose as free parameters $H_\mathrm{DMI}, \alpha_0, H_K$, where we allow the anisotropy field to vary with respect to the measurement: $H_K=2 K / \mu_0 M_s \pm s_{H_K}/2$, with\\ $s_{H_K}^2=H_K^2\left(s_K^2/K^2+s_{M_s}^2/M_s^2 \right)$. To apply Eq.\ref{eq:vbase} we need the values of the characteristic speed $v_0$ and the scaling constant $\alpha_0$. These are obtained by the preliminary $v(H_z)$ creep measurement shown in Fig.\ref{fig:creepbehaviour}.
The values of the physical parameters $t_i$, $M_s$, $H_K$, $A$ are obtained by independent measurements with a given uncertainty as described previously.
The fit then returns the optimal parameter values and their estimated covariance.

\begin{figure}    
 \centering
    \includegraphics[width=\linewidth]{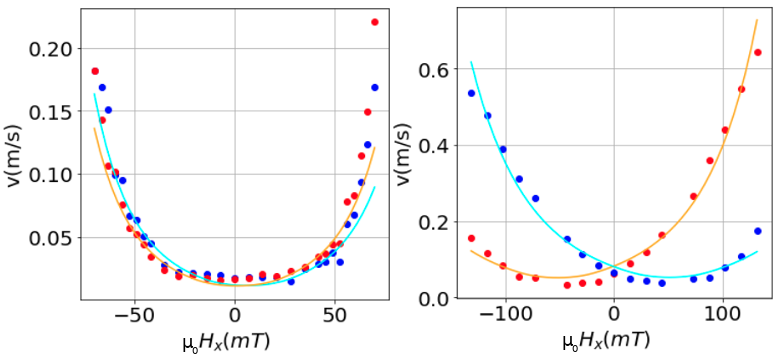}
    \caption{W/CoFeB based samples: velocities $v_{\uparrow \downarrow}$ (red dots) and  $v_{\downarrow \uparrow }$ (blue dots), and the standard creep model fits (lines) from Eq.\ref{eq:vbase}, for as-deposited \textit{756b} sample (left), annealed sample \textit{756a} sample (right)}
    \label{fig:JeFitWCoFeB}
\end{figure}
\begin{figure}    
 \centering
    \includegraphics[width=\linewidth]{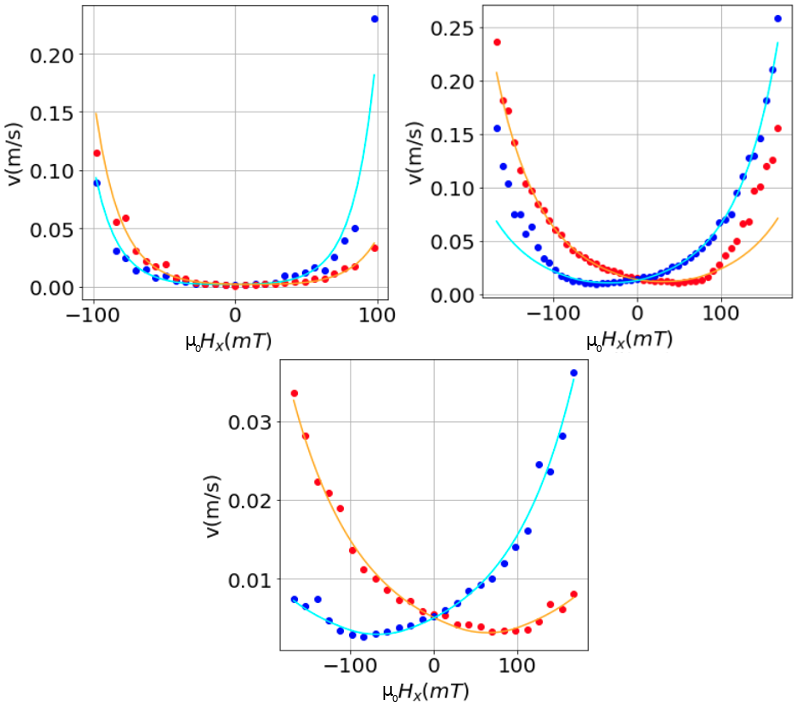}
    \caption{Co-based samples: velocities of $v_{\uparrow \downarrow}$ (red dots) and $v_{\downarrow \uparrow }$ (blue dots), and the standard creep model fits (lines) from Eq.\ref{eq:vbase}, for \textit{a1} sample (top left), \textit{a4} sample (top right), \textit{a5} sample (bottom)}
    \label{fig:JeFitLeeds}
\end{figure}

In the samples where regular bubbles can be found, the standard creep model can be applied (e.g. W/CoFeB samples Fig.\ref{fig:JeFitWCoFeB}, Co samples Fig.\ref{fig:JeFitLeeds}), even when analyzing rough bubble domains. The exceptions are the irradiated samples, which are characterized by a particularly asymmetric DW velocity curve.
The standard model is not perfect however. 
Even in the presence of smooth, isolated bubbles, the model in some cases is not correct in describing the domain wall expansion. 
The shape of velocity curves can be asymmetric with respect to the $H^0_x$ minimum (e.g. Fig.\ref{fig:JeFitLeeds}, top right), or, especially for irradiated samples, a clear change of slope can be observed at higher fields. As we mentioned, it is possible to address this problem by some extended models (as discussed in the following).
Another difficulty is that the velocity curves can be anomalously flat (e.g. Fig.\ref{fig:JeFitLeeds}, top left)), deviating from the expected quadratic behavior in $H_x$.
Finally, we can see that the model is strongly sensitive to the value of some physical parameters. In particular, a good fit often requires a low value for the anisotropy field $H_K$, beyond the error threshold. This is consistent with the fact that the asymmetric bubble expansion method is based on the behaviour of the DW both in the hybrid Bloch/Néel state (at low velocity, Eq.~\ref{eq:sigmaDW2}) and in the Néel state (at high velocity, Eq.~\ref{eq:sigmaDW1}), and it has been observed \cite{BER-92,BER-93} in Co thin films that the Néel wall profile along its thickness is correctly obtained only with an anisotropy constant substantially lower than the value measured by other experimental techniques.

\begin{figure}    
 \centering
    \includegraphics[width=0.8\linewidth]{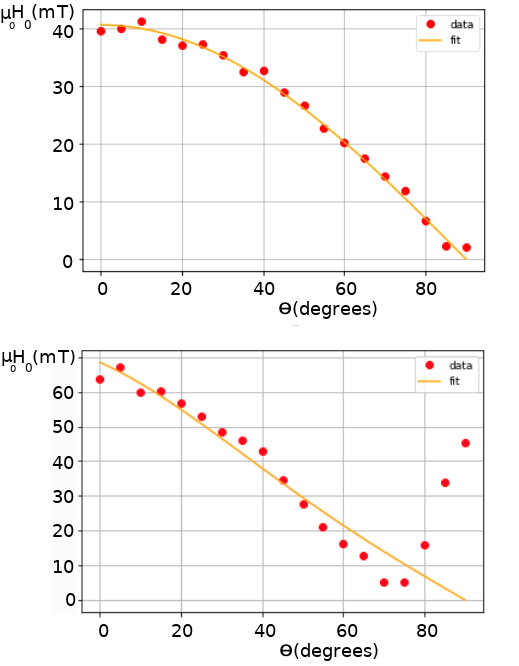}
    \caption{$H_0$ measurements (dots) and fit (line) by Eq.~\ref{eq:H0_arbangle} of W/CoFeB \textit{756a} sample (top): $\mu_0 H_\mathrm{DMI} = 40.6$ mT, $\mu_0 H_\mathrm{K}^\mathrm{DW} = 0.$ mT textcolor{red}{missing value after decimal point?} and Co-based sample \textit{a5} (bottom): $\mu_0 H_\mathrm{DMI} = 68.5$ mT, $\mu_0 H_\mathrm{K}^\mathrm{DW} = 29.4$ mT}
    \label{fig:arbanglefit}
\end{figure}

The arbitrary angle propagation model is applied by using Eq.~\ref{eq:H0_arbangle} to identify the DMI field.
This is accomplished by exploring the absolute value of the position of the minima of the curve velocities at each direction in the range 0 to $\pi/2$.
The application of Eq.\ref{eq:H0_arbangle} then allows us to obtain the value of $H_\mathrm{DMI}$, and also an independent estimate of $H_\mathrm{DW}$.
As in the other methods, in the presence of asymmetry effects in the velocity curves, an error can be introduced in the determination of $H_\mathrm{DMI}$. 
In \cite{KIM-17} the method was applied to a straight DW, so in our application of this model we nucleated large magnetic bubbles, to reduce the DW curvature.
It is an important result that we find the model can in fact be applied to circular bubbles, as shown for the sample \textit{756a}, Fig.~\ref{fig:arbanglefit} (top), where the bubble expansion correctly follows Eq.~\ref{eq:H0_arbangle} and allows the identification of $H_\mathrm{DMI}$. When the bubble expansion is influenced by neighboring bubbles though, as in sample \textit{a5}, Fig.~\ref{fig:arbanglefit} (bottom), it is not always possible to obtain a good result in arbitrary samples. This method is particularly sensitive both to bubble deviations from circularity, and to the proximity to other bubbles: in this last case, a strong deformation appears beyond $60^\circ$, making the model fail.

\begin{figure}    
 \centering
    \includegraphics[scale=0.60]{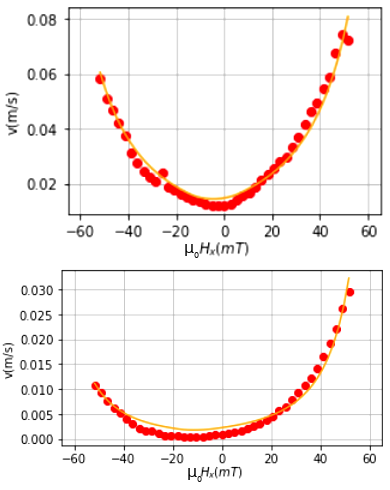}
    \caption{\textit{ID0} (top) and \textit{ID16} sample (bottom); measurements of $v_{\uparrow \downarrow}$ (dots) and fit (line) by the varying DW width model
    }
    \label{fig:varyingwidthfit}
\end{figure}

In the varying DW width model Eqs.~\ref{eq:sigmaeqvaryingwidth1}, \ref{eq:sigmaeqvaryingwidth2} one makes use of the new definition for the DW energy density in Eq.~\ref{eq:vbase}.
Although it has been mentioned that the DW energy density becomes asymmetric in $H_x$, it can be demonstrated that the maximum in $\sigma$ remains at the same value independently of $D$, and that at the field value where $\sigma$ is maximum the DW is Bloch-like. Therefore, the relationship $H_\mathrm{DMI}=H_0= -D / \mu_0 M_\mathrm{s} \lambda$, where $H_0$ is the velocity curve minimum position, remains valid.
One of the features of this model is that the asymmetry decreases with decreasing $D$, with the model tending to the limit of the standard creep model. Yet in some samples we find strongly asymmetric curves even at rather low $|D|$ values.
The only exception is found in the irradiated samples, with two cases shown in Fig.~\ref{fig:varyingwidthfit}, showing that this variation of the standard model is able to correctly reproduce asymmetric velocity curves in selected cases.

\begin{figure}    
 \centering
    \includegraphics[width=\linewidth]{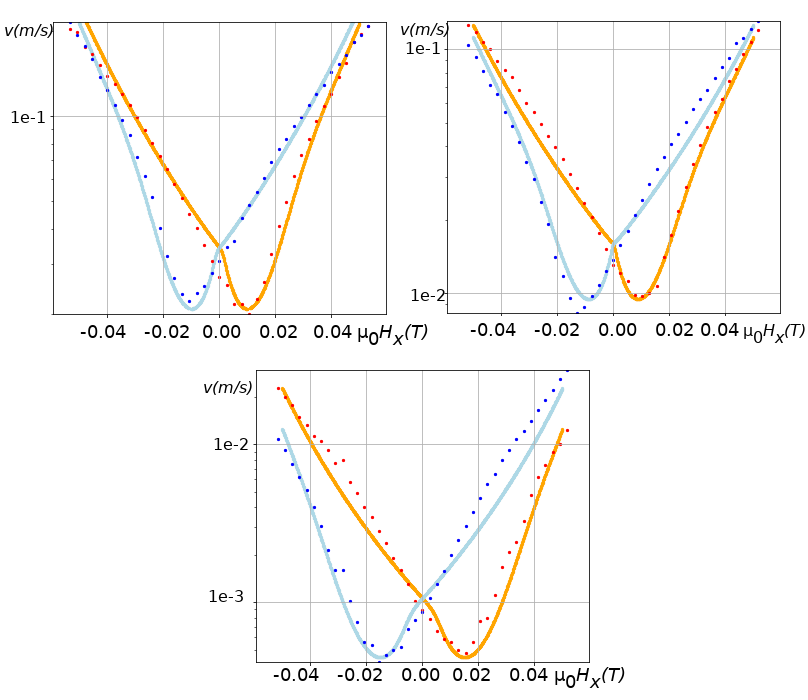}
    \caption{Irradiated samples \textit{ID8} (top left), \textit{ID12} (top right) and \textit{ID16} (bottom): velocities $v_{\uparrow \downarrow}$ (cyan dots) and $v_{\downarrow \uparrow }$ (orange dots), and dispersive stiffness model fits (lines).
    }
    \label{fig:dispersivefit}
\end{figure}

In the dispersive stiffness model, the stiffness is used to describe the presence of a highly anisotropic interface energy in the DW, by using Eq.~\ref{eq:sigmastiff} in Eq.~\ref{eq:vbase}. After finding the $\psi_{eq}$ value such that $\sigma$ is minimum, the $\sigma_{\theta\theta}$, $\sigma_{\theta\psi}$, $\sigma_{\psi\psi}$ double derivatives are calculated. These derivatives are necessary to obtain the second order expansion of the energy density about $\theta_0$ and $\psi_0$ for a straight domain wall segment.
The model includes an additional parameter $L$ describing the DW deformation length scale. It can exhibit large values for low-coercivity materials with a sparse distribution of pinning sites, or it can be set to $L\rightarrow0$ for sputtered thin films which exhibit a dense distribution of pinning sites. In any case, for small $L$, the results are not very sensitive to $L$.
The velocity described by Eq.~\ref{eq:stiffvelocity} is successful in describing the different slopes of the velocity curves in the irradiated samples, particularly in high $|D|$ cases, Fig.~\ref{fig:dispersivefit}. The absolute value of the DMI value obtained by this method is found to be between $15\%$ and $20\%$ larger, when compared with that obtained by the parabolic fitting of the minimum.
While in the application of the standard creep model the fit was performed by a non-linear least squares procedure, the higher complexity of the expression Eq.~\ref{eq:stiffvelocity} forced us to use a manual fit, with the results shown in  Fig.~\ref{fig:dispersivefit} for samples listed in Tab.~\ref{table:implanted samples}). 

Tables Tab.~\ref{table:Co D} and Tab.~\ref{table:CoFeB D} show the results of the Co- and CoFeB-based samples measured at Leeds and INRIM using a parabolic fit and the standard creep model. In Tab.~\ref{table:irr D} the $D$ values obtained for the irradiated samples from the dispersive stiffness model are shown. For the Co-based samples we find reasonable measurement errors from the fit of about 5-10\% of the measured value (except when the $D_\mathrm{s} = D t_\mathrm{FM}$ value is very small, i.e. of the order of 0.01~pJ/m). For the CoFeB-based samples we find very different measurement errors. In most cases the errors are of the order of 0.01~pJ/m, as for the Co-based samples, but errors as small as 0.002~pJ/m (756a) and bigger than the measured value itself (758a) are obtained. The latter sample is characterised by a rough bubble DW, with a noisy velocity curve. The error can be reduced by using the more accurate standard creep model: however it remains substantial for the measurement performed at Leeds due to the slightly more scattered data and smaller applied IP field range.

The irradiated samples' results are shown in Tab.~\ref{table:irr D}. The bubble domain walls are smooth at all irradiation values, with occasionally some pinning lines interfering in the velocity evaluation. The velocity curves have a high curvature around the minimum, although they are strongly asymmetrical, as mentioned.
The quality of the bubble domain and of the velocity curves changes substantially across the sample series and leads to different applicability of the standard creep model. 

\begin{table*}
\caption{Co samples with Pt and/or Ir heavy metal layer prepared at University of Leeds. The $H_\mathrm{DMI}$ values obtained from the data measured at Leeds by using a parabolic fit and from data measured at INRIM, using the standard creep model are compared. The $D$ values were obtained from the INRIM data by the standard creep model, $K$ and $\alpha_0$ are fitting parameters. $D_\mathrm{s}$ is calculated from $D$ by $D_\mathrm{s}=D t_\mathrm{FM}$, where $t_\mathrm{FM}$ is the nominal thickness of the FM film. The value of $M_\mathrm{s}$ was obtained by the `original' method, using the nominal Co thickness.
}
\small
\begin{tabular}{c|c|c|c|c|c|c|c|c|c|c}
\hline
\hline
\textbf{\scriptsize{sample}}&\textbf{$\mu_0 H_\mathrm{DMI}^\mathrm{Leeds}$}&\textbf{$\mu_0 H_\mathrm{DMI}^\mathrm{INRIM}$}&\textbf{$M_\mathrm{s}$}&\textbf{$K_\mathrm{eff}$}&\textbf{$A$}&\textbf{$\lambda$}&\textbf{$K$}&\textbf{$\alpha_0$}& \textbf{$D$}&\textbf{$D_\mathrm{s}$}\\
\hline
&\scriptsize{(mT)}&\scriptsize{(mT)}&\scriptsize{(MA/m)}&\scriptsize{(MJ/m$^3$)}&\scriptsize{(pJ/m)}& \scriptsize{(nm)}&\scriptsize{(MJ/m$^3$)}&\scriptsize{($T^{1/4}$)}   &\scriptsize{(mJ/m$^2$)}&\scriptsize{(pJ/m)}\\
\hline
\hline
a1&-5.7$\pm$0.4& -0.6$\pm$2.0  &1.58$\pm$0.03& 0.540$\pm$0.02 & 12.6$\pm$2 & 4.8$\pm$0.4  & 0.610$\pm$0.02 & 26.16$\pm$0.01 & 0.11$\pm$0.01& 0.09$\pm$0.01\\
a2&-2.3$\pm$0.2& -2.0$\pm$2.0&1.74$\pm$0.04& 0.470$\pm$0.02 & 14.1$\pm$2 & 5.5$\pm$ 0.8     &  0.420$\pm$0.03 & 22.87$\pm$0.01 & 0.14$\pm$0.01&0.11$\pm$0.01\\
a3&36.7$\pm$0.9& 17$\pm$5 &1.14$\pm$0.05& 0.430$\pm$0.02 & 7.1$\pm$2 & 4.1$\pm$ 1.0      &  0.441$\pm$0.001 & 17.40$\pm$0.02 & -0.05$\pm$0.01&-0.04$\pm$0.01\\
a4&51$\pm$3& 12$\pm$1 & 1.20$\pm$0.09& 0.420$\pm$0.04 & 5.8 $\pm$2 & 3.7$\pm$ 1.0   &0.751$\pm$0.002 & 12.87$\pm$0.01 & -0.25$\pm$0.02& -0.20$\pm$0.02\\
a5&78.7$\pm$0.8& 36$\pm$3 &0.90$\pm$0.02& 0.570$\pm$0.02 & 7.6 $\pm$2 & 3.7$\pm$1.0 & 0.897$\pm$0.005 & 13.42$\pm$0.02 & -0.36$\pm$0.02&-0.29$\pm$0.02\\
\hline
\hline
\end{tabular}

\label{table:Co D}
\end{table*}


\begin{table*}
\caption{CoFeB-based samples with W (756a/b) or Pt (758a to 762a) heavy metal layer prepared at University of Mainz. The $H_\mathrm{DMI}$ values obtained from the data measured at Leeds by using a parabolic fit and from data measured at INRIM, using the standard creep model are compared. The $D$ values were obtained by the standard creep model on the data measured at INRIM. The CoFeB-based samples with several repetitions of the FM/HM layer are not included since it was not possible to determine $D$.\\$^{\mathrm{\textcolor{blue}{1}}}$This is a typical value extracted from the literature. Temperature dependent measurements on these samples were not performed.}
\small
\begin{tabular}{c|c|c|c|c|c|c|c|c|c|c}
\hline
\hline
\textbf{\scriptsize{sample}}& \textbf{$\mu_0 H_\mathrm{DMI}^\mathrm{Leeds}$} &\textbf{$\mu_0 H_\mathrm{DMI}^\mathrm{INRIM}$}&$M_\mathrm{s}$& \textbf{$K_{\mathrm{eff}}$}&$A$\footnote{}&$\lambda$&\textbf{$K$}&\textbf{$\alpha_0$}& \textbf{$D$}&\textbf{$D_\mathrm{s}$}\\
\hline
& \scriptsize{(mT)}& \scriptsize{(mT)} &\scriptsize{(MA/m)}&\scriptsize{(MJ/m$^3$)} &\scriptsize{(pJ/m)}&\scriptsize{(nm)}& \scriptsize{(MJ/m$^3$)} & \scriptsize{($T^{1/4}$)}   &\scriptsize{(mJ/m$^2$)}&\scriptsize{(pJ/m)}\\
\hline
\hline
756a &-45.2$\pm$6&  -52.0$\pm$5  & 0.92$\pm$0.06& 0.730$\pm$0.06&20&5.4$\pm$0.2& 0.519$\pm$0.003 & 6.15$\pm$0.01 & 0.237$\pm$0.003&0.142$\pm$0.002\\
756b & - & -4.0$\pm$0.1 & 1.65$\pm$0.2&- &-&- & 0.195$\pm$0.003 & 6.80$\pm$0.04 &  0.021$\pm$0.009&0.013$\pm$0.005\\
\hline
758a &2.5$\pm$5& 9.0$\pm$0.7 &1.56$\pm$0.04& 0.800$\pm$0.06 &20& 4.5$\pm$0.5 & 0.441$\pm$0.003 & 12.42$\pm$0.09 & -0.1$\pm$0.1 &-0.1$\pm$0.1\\
760a &5.0$\pm$4& 13.0$\pm$0.6   &1.58$\pm$0.04& 1.470$\pm$0.07&20&3.7$\pm$0.1 &0.504$\pm$0.003 & 10.90$\pm$0.06 &-0.14$\pm$0.03&-0.11$\pm$0.02 \\
762a &12.6$\pm$3& 20.0$\pm$0.2 &1.62$\pm$0.05& 1.450$\pm$0.08 &20&3.7$\pm$0.1& 0.541$\pm$0.001 & 14.33$\pm$0.02 & -0.33$\pm$0.06 &-0.27$\pm$0.05\\
\hline
\hline
\end{tabular}
\label{table:CoFeB D}
\end{table*}

\begin{table*}
\caption{He-irradiated CoFeB-samples. The $D$ values were obtained by the dispersive stiffness model. The error in $\alpha_0$ is obtained by the creep law fit. }
\small
\begin{tabular}{c|c|c|c|c|c|c}
\hline
\hline
\textbf{\scriptsize{sample}}& \textbf{\scriptsize{$\mu_0 H_\mathrm{DMI}$}} & \textbf{\scriptsize{$M_\mathrm{s}$}}& \textbf{\scriptsize{$K$}} & \textbf{\scriptsize{$\alpha_0$}} &\textbf{\scriptsize{$D$}}&\textbf{\scriptsize{$D_\mathrm{s}$}}\\
\hline
& \scriptsize{(mT)} & \scriptsize{(MA/m)}& \scriptsize{(MJ/m$^3$)} & \scriptsize{($T^{1/4}$)}   &\scriptsize{(mJ/m$^2$)}&\scriptsize{(pJ/m)}\\
\hline
\hline
\scriptsize{ID0}& 3$ \pm$ 5&$0.87\pm0.09$& $0.120\pm0.006$ & $3.16\pm0.05$ & $0.06\pm0.03$ & $0.06\pm0.03$ \\
\scriptsize{ID4}& 10$\pm$2 &$0.75\pm0.08$& $0.095\pm0.005$& $3.0\pm0.2$ & $0.11\pm0.01$ & $0.11\pm0.01$ \\
\scriptsize{ID8}& 13$\pm$3 &$0.83\pm0.08$& $0.096\pm0.005$ & $3.00\pm0.08$ & $0.19\pm0.01$ & $0.19\pm0.01$\\ 
\scriptsize{ID12}& 15$\pm$3 &$0.71\pm0.07$& $0.094\pm0.004$ & $3.2\pm0.1$ & $0.19\pm0.01$ & $0.19\pm0.01$ \\
\scriptsize{ID16}& 18$\pm$4 &$0.65\pm0.07$&$0.094\pm0.004$ & $3.56\pm0.07$ & $0.27\pm0.02$ & $0.27\pm0.02$ \\
\hline
\hline
\end{tabular}
\label{table:irr D}
\end{table*}

\subsection{Reliability of results and comparison with BLS}

The results were compared in an international ``round robin'' (RR) effort to determine the DMI \cite{TOPS}. The participating laboratories determined the DMI with their preferred procedure on a coupon from the same wafer to avoid any sample-to-sample variations. INRIM used the standard creep law for fitting the velocity data, while the University of Leeds employed parabolic fitting for obtaining the field $H_{\mathrm{DMI}}$. As shown in Tab.~\ref{table:Co D}, for the Co-based samples the values differ significantly, although the measurement error in most cases is reasonably small. For the CoFeB-based samples the measurement error is significantly larger but the overall agreement between the results obtained in the two laboratories is slightly improved. Possible sources for these deviations are a) inhomogeneities of the sample, b) systematic errors due to the different evaluation of the minimum of the DW velocity curve, c) difficulties in fitting due to noise in the velocity curves or insufficient IP field range. 

\begin{figure}    
 \centering
    \includegraphics[width=\linewidth]{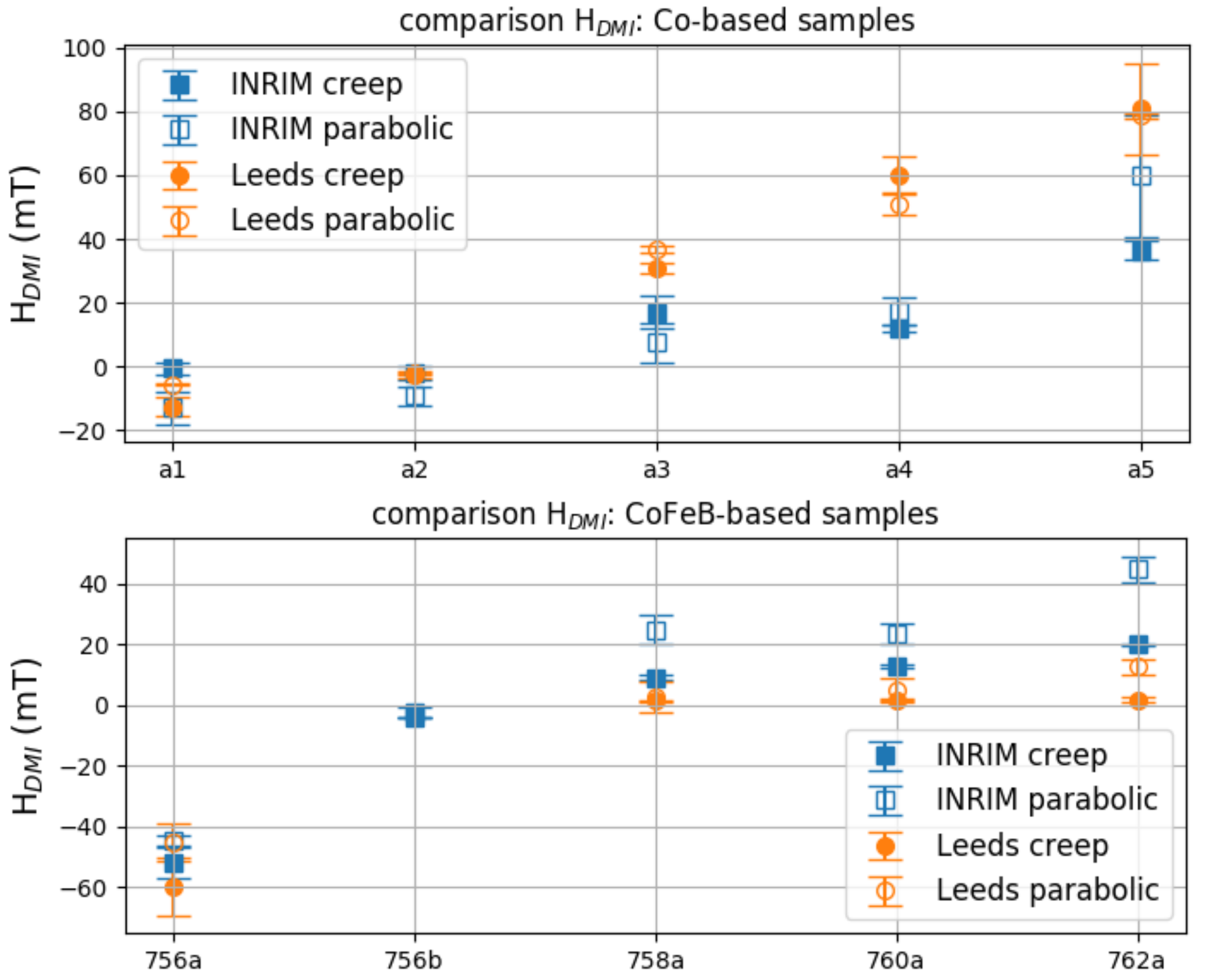}
    \caption{Comparison of the RR results of the Co-based (upper panel) and CoFeB-based (lower panel) samples. For the latter the data measured at Leeds were fitted by a parabola, according to their standard procedure, as well as by using the standard creep model. The data measured at INRIM were fitted by the standard creep model.
    }
    \label{fig:RRHdmi}
\end{figure}

In order to see if the evaluation by the creep model with respect to the parabolic fit changes strongly the results, or if the discrepancies are due to differences in the data, the data sets of both laboratories were evaluated by both models. The result is shown in Fig.~\ref{fig:RRHdmi}. The systematically lower values for the Co-based samples, and higher values for the CoFeB-based samples, measured at INRIM, point in the direction that Statistical inhomogeneities play a minor role. While for the Pt/CoFeB-based samples (758a, 760a and 762a) the measurement error could be reduced by using the standard creep model with respect to a simple parabolic fit, the $H_\mathrm{DMI}$ values show even larger deviations from the ones measured at INRIM. For the W/FeCoB (samples 756a and 756b ) and the Co based samples (a1-a5) the agreement was neither improved nor worsened. We note also that for the Leeds' measurements the model seems to play a minor role, while for INRIM's data the deviation between the $H_{DMI}$ values obtained by the two different models is large. We find differences of 2-3 times the measured value (e.g. samples a5 and 762a). A reason for the discrepancy, independently of the model used, may be that the measurements performed at Leeds are limited to smaller IP fields. This means that the fitting range is reduced which may result in a shift of the velocity curve minimum, especially in the case of asymmetric velocity curves. This also may explain the smaller difference for the two applied models, since the asymmetry shows up at higher IP fields. Another difference between the two laboratories is that the measurements at INRIM were performed at higher OOP field values and the velocities are orders of magnitude higher.  The lower velocities may also lead to slightly more noisy curves, since pinning plays a stronger role. This was observed especially for samples with poor repeatability (as indicated in Fig.~\ref{fig:WCoFeB}).

\begin{figure}    
 \centering
    \includegraphics[width=\linewidth]{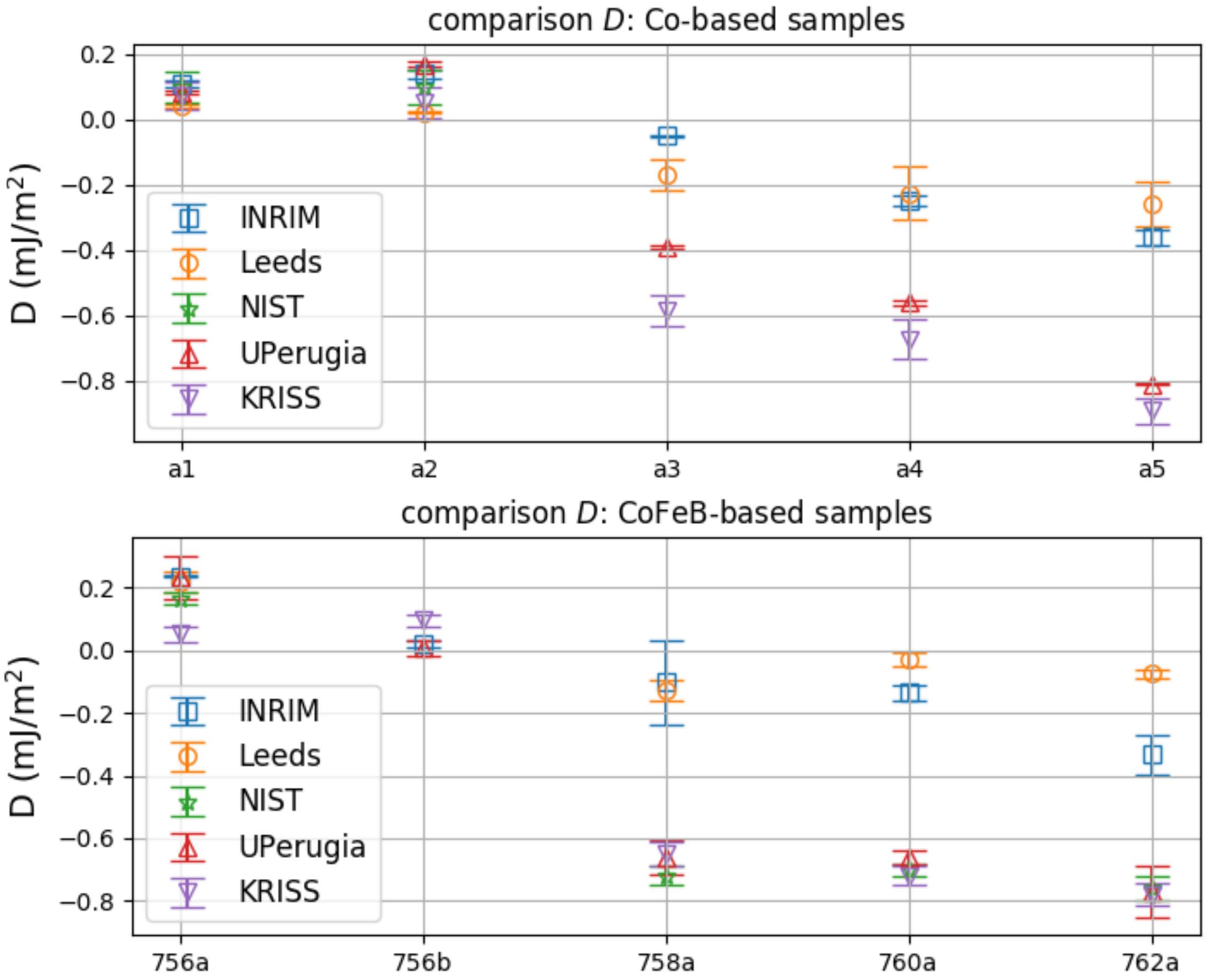}
    \caption{Comparison of the RR results of the Co-based (upper panel) and the CoFeB-based (lower panel) samples. UPerugia, KRISS and NIST extracted $D$ from the spin wave non-reciprocity measured by Brillouin light scattering.
    }
    \label{fig:RRD}
\end{figure}

The $D$ value was obtained from Eq.~\ref{eq:HDMI}, with the $D$ error obtained from propagation of the fitting error of $H_\mathrm{DMI}$, $M_\mathrm{s}$ and $\lambda$. The results were compared with independent measurements of the DMI value performed by Brillouin light scattering (BLS) performed at the University of Perugia (UPerugia), the Korea Research Institute of Standards and Science (KRISS) and  National Institute of Standards and Technology (NIST). The results are shown in Fig.\ref{fig:RRD}. The agreement for the Co-based samples between BLS and MOKE, considering the error bars, is generally better than the agreement for the CoFeB based samples, where the reported values for BLS are much larger than for MOKE for three out of five samples.

Disagreements between BLS and MOKE based methods have been reported in the literature \cite{LAU-18,DIE-19}. It was suggested that creep models can fail in the limit of low damping, where features like roughening of the domain walls can appear in the domain structures \cite{SOU-16}.
Also, it was discussed in \cite{KIM-19} that differences between BLS and domain expansion methods could be due to asymmetries in the velocity curves measured by the latter, asymmetries that are found to increase with the ferromagnetic layer thickness. 

Furthermore, the two methods may be sensitive to defects at length scales differing orders of magnitude, since BLS averages the result over a spot size of $\approx 10-40~\upmu$m, revealing thermally excited spin waves with  wavelengths of about half a $\upmu$m, while the bubble expansion method, based on creep DW motion, probes the system at the nanoscale. In fact, in the creep regime a balance is established between the pinning effect of defects interacting with the DW, and the DW elastic energy. And the defects distribution is found at length scales of the order of the DW width $\lambda \approx 10$~nm \cite{SOU-16}. As an example, in a Pt/Co(0.5-0.8)/Pt system the pinning length above which the DW elastically adjusts to a local energy minimum is around 25 nm \cite{LEM-98}.

\subsubsection{Repeatability of the measurement}

We compared several measurements performed on the same sample either of the same bubble at different OOP fields or at different bubbles and find in some cases substantial deviations. For the CoFeB-based samples we find a reasonably good repeatability of the measurement for the sample with W (756a), which has a high $H_\mathrm{DMI}$. Instead, for the Pt samples differences in $H_\mathrm{DMI}$ may be up to 50\%, as shown by the large errors in Table \ref{table:CoFeB D}. This might be related to problems of repeatability in the bubble nucleation but also due to bad fits caused by flat or asymmetric velocity curves. Sometimes the bubble nucleated before the asymmetrical expansion is not reappearing at the same position, or it assumes significantly different shapes and sizes despite the use of the same OOP nucleation field and site for each image. Reasons for the velocity noise besides the bubble nucleation may be the difficulty to determine reliably the DW velocity due to rough DWs. Having complete, high-quality velocity curves available and reliable methods for determining the DW velocity, such as MOKAS, are therefore very important to reduce fitting errors. Furthermore, measurements have to be repeated several times in order to have sufficient statistics to define the confidence interval. 

\subsubsection{Differences in results from calibration errors}
The main uncertainties in the measurement chain relate to the error in the measurements of the fields, and in the calculation of the domain wall velocities. The relative uncertainty in the in-plane field measurement is $\delta H_x/H_x=3 \cdot 10^{-3}$, while for the OOP field we have $\delta H_z/H_z=1.5 \cdot 10^{-2}$. 
The error in the velocities calculation is $\delta v /v=2 \cdot 10^{-3}$, estimated from the error in the spatial position of the wall (depending on the magnification) and to the error in the time measurement.


The DMI value $D$ was calculated by $D=-\mu_0 M_\mathrm{s} \lambda H_\mathrm{DMI}$, with a relative uncertainty $s_D^2/D^2=s_{M_\mathrm{s}}^2/M_\mathrm{s}^2+s_{H_\mathrm{DMI}}^2/H_\mathrm{DMI}^2+s_{\lambda}^2/{\lambda}^2$, where $s_{\lambda}=\lambda\sqrt{s_A^2/4A^2+s_K^2/4K^2}$.

\section{Conclusions}

The asymmetric bubble expansion method can be applied to a wide variety of heterostructures with perpendicular magnetic anisotropy where the domains are bubble-like. The evaluation of the DMI value $D$ is performed in two steps: 1) extracting the DW velocities under applied in-plane field from the expanding bubble by magneto-optical imaging, 2) evaluating $D$ from the velocity curves by using appropriate models. The accuracy of the result depends critically on both steps. Furthermore, the applicability of the method is limited by interface quality, which is related to the bubble DW roughness. In extreme cases, such as in multilayers with several repetitions the interface quality may deteriorate and we observed rough DWs or even transitions to maze domains, so that the method cannot be applied. We summarize here what are considered the ``key points'' for a reliable determination of $D$ by this method and give indications on how to reduce the measurement uncertainty.\\

\emph{The applied magnetic field}\\
Since the method is based on the compensation of the DMI field $H_{\mathrm{DMI}}$ by an applied in-plane field, the uncertainty in the applied field is the main source for uncertainty in $D$. We therefore recommend to calibrate carefully the field and check for any tilt of the sample which will contribute to the OOP component of the field. Furthermore, $H_z$ has to be chosen well inside the creep regime, and constant pulse duration must be kept in all measurements. \\

\emph{Acquisition of the bubble expansion velocity}\\
The acquisition of the bubble expansion velocity can be problematic if the bubble DW is rough or if there are problems with repeatable nucleation of the bubble, as we found for the W/CoFeB samples. We therefore recommend to check for the repeatability and average over several measurements. The nucleation is influenced by inhomogeneities and some bubbles show different expansions which result in $D$ values up to 50\% different due to the altered local energy landscape. Furthermore, the expansion is changed also in presence of other bubbles close by. We find a correlation between DW roughness and top layer roughness. The rougher the bubble DW, the more difficult is it to identify the DW velocity and its minimum. For rough bubbles it might be convenient to try the MOKAS software, openly available, which offers an automatic evaluation of the velocity curve.\\
 
\emph{Extraction of the velocity curves and modelling of the $D$ value}\\
To extract $D$ from the velocity curve, the choice of the model and the fitting procedure are most critical. Choosing different models to fit the same data, as we did by comparing a parabolic fit with the standard creep model, results in $D$ values varying about to 2-3 times. As a basic recommendation we suggest to extend the $H_x$ range to the maximum, to ensure a more accurate fit.

Which model has to be used depends upon the shape of the velocity curve, which is correlated to the DW roughness and the interface and sample quality. Rougher DW lead to flatter velocity curves (or curves that deviate more from a parabola). Such curves cannot be fitted by the standard creep theory which predicts quadratic growth of the DW velocity around the minimum. The arbitrary angle propagation model is useful if the direction of fastest expansion is not well defined, and is very sensitive to the presence of magnetostatic interaction from neighboring bubbles. Samples with induced defects by irradiation have bubble DWs slightly rough, but well defined and the velocity curves are well determined. There is no flat region but a change of slope well described by the dispersive stiffness model. In general, it is difficult to give a recipe which model applies to which kind of sample. The shape of the velocity curve has to be examined. However, while defects and pinning increase DMI, they limit the applicability of this technique, regardless of the model used.\\

\emph{Determination of the model parameters}\\
The choice of the material parameters which enter in the models ($M_\mathrm{s}$, $A$ and $K_\mathrm{eff}$) and their uncertainty reflects directly in the uncertainty of $D$. We recommend therefore to measure and evaluate carefully these parameters. The determination of $M_\mathrm{s}$ is complicated by the presence of dead or proximity layers. Furthermore, the fitting result of the experimental data to a given model is strongly sensitive to small variations of several of the physical parameters, in particular $M_\mathrm{s}$ and $K_\mathrm{eff}$. We also verified that in many cases a good fitting of the data using the chosen model requires the setting of the anisotropy constant much lower than the measured value, beyond the error threshold.\\

Considering these indications, the asymmetric bubble expansion method is able to deliver reliable results. However, it remains to be investigated how the results compare with different methods. Cross-checking with BLS measurements on the same samples, we find systematic discrepancies. While for small $D$ values the agreement is within the measurement error, at higher $D$ we find systematically higher values of $|D_{BLS}|$ with respect to $|D_{mopt}|$. The large discrepancies in some cases cannot be completely eliminated by a better choice of the model applied to the asymmetric bubble expansion data. We suggest therefore that this can be due to the sensitivity to local $nm$-scale defects or inhomogeneities, not observed in the micrometer-scale averaged BLS measurements, an issue worth to be investigated in future works.




\section*{Data and code availability}
The data and the code associated with this paper are openly available from \doi{10.5281/zenodo.5844251} .\\
The software \emph{MOKAS} is available at GitHub,\\ \doi{10.5281/zenodo.5714377}, \href{https://github.com/gdurin/mokas}{MOKAS software}.

\section*{Acknowledgements}
The project 17FUN08-TOPS has received funding from the EMPIR  programme  co-financed  by  the  Participating  States and from the European Union’s Horizon 2020 research and innovation programme. The group in Mainz acknowledges support from the German Research Foundation (SFB TRR 173 Spin+X No. 268565370, projects A01 and B02 and No. 403502522-SPP 2137 Skyrmionics) and from the Horizon 2020 framework program of the European commission under grant No. 856538 (ERC-SyG 3DMAGIC). HTN and JMS acknowledge support by the DARPA Topological Excitations in Electronics (TEE) program, award No. R18-687-004.

\appendix
\section{Influence of errors in the determination of the material parameters on the uncertainty of $D$} \label{sec:app}

Since the material parameters $M_\mathrm{s}$, $K_{\mathrm{eff}}$ and $A$ are required for the evaluation of $D$, any error in their determination will directly reflect in the error of the DMI value. 

In particular, the measurement of $M_\mathrm{s}$ requires a careful analysis of the magnetometry data including the sample area and the thickness. For each of the Co-based samples the magnetic moment $m$ was obtained from a fit to the data measured by SQUID and the saturation magnetization was calculated by $M_\mathrm{s}=m/V$, where $V$ is the volume of magnetic material in the sample. The thickness $t$ was measured by low-angle X-ray reflectivity spectra. We find actual Co thicknesses of 1.01~nm to 1.07~nm ($\pm$0.01~nm) from these fits, with respect to the nominal one of 0.8~nm. The GenX code \cite{Bjorck:aj5091} was used to determine not only a sample's thickness, but also density and roughness from the reflectivity fits. Each sample area $a$ was calculated from an image of the sample together with a scale. The images were taken directly after the SQUID measurements. The number of pixels per millimeter was determined from the scale and the number of pixels within the sample area was extracted by manually defining the sample area in the image. This process was averaged three times and the error was set as half of the difference between the smallest and largest measured sample areas. Standard error propagation was then used to determine the error in $M_\mathrm{s}$ from the error in $t$ and $a$.

However, there are additional problems in the evaluation since the correct ``magnetic volume'' is often not known or ill-defined, due to the presence of dead or proximity layers when the FM layer is in close contact to a HM layer \cite{BAN-11}. Using the measured Co thickness for calculating the "magnetic volume" we notice large deviations in $M_s$, see Table~\ref{table:Co D} (from 0.9~MA/m to 1.74~MA/m) although all samples are composed of the same nominal Co film, with relative errors in the Co thickness of maximum 5\%. 

In order to investigate how these interfacial effects affect the measured magnetic moment, we grew for each material combination shown in Tab.~\ref{table:Co samples} different thicknesses of Co (1.5~nm, 4.0~nm, 8.5~nm, and 17~nm) sandwiched between two layers of either Pt, Ir, or Ta. For each material combination we then plotted the moment/area against Co thickness and determined the y-axis intercept, which is negative for Ta and positive for Pt and Ir\footnote{The plots are not shown here for the sake of being more concise.}. This indicates the presence of a magnetic dead layer at the Ta interface and induced magnetic moments for Ir and Pt. For the case of Ta, the x-axis intercept represents the thickness of the dead layer, whereas for the Pt and Ir, it represents the extra thickness of Co that would be required to make up for the increased in moment at the interface. With this information, we applied two different methods to account for the dead layer/induced moment in the samples in order to get a more accurate value for $M_\mathrm{s}$. The first method, defined as the ``moment/area method'', involved subtracting or adding the moment/area for each interface to obtain a moment/area for solely the Co. The second method, defined as the ``thickness method'', involved subtracting/adding the thickness gained/lost so that the thickness used in the $M_\mathrm{s}$ calculation corresponded to that of all the present moments. The ``original'' method used only the Co parameters and did not consider any interfacial effects to define $M_\mathrm{s}$. The lack of consideration of dead layers or proximity magnetization can result in errors of up to 66\% in $M_\mathrm{s}$. After application of both methods, the $M_\mathrm{s}$ values are more consistent among the five samples, averaging $0.7 \pm 0.1$~MA/m for the moment/area method and $0.9 \pm 0.2$~MA/m for the thickness method: values more reasonable for a magnetic Co thin film \cite{MET-07}. 

\begin{figure}[t]    
 \centering
   \includegraphics[width=0.8\linewidth]{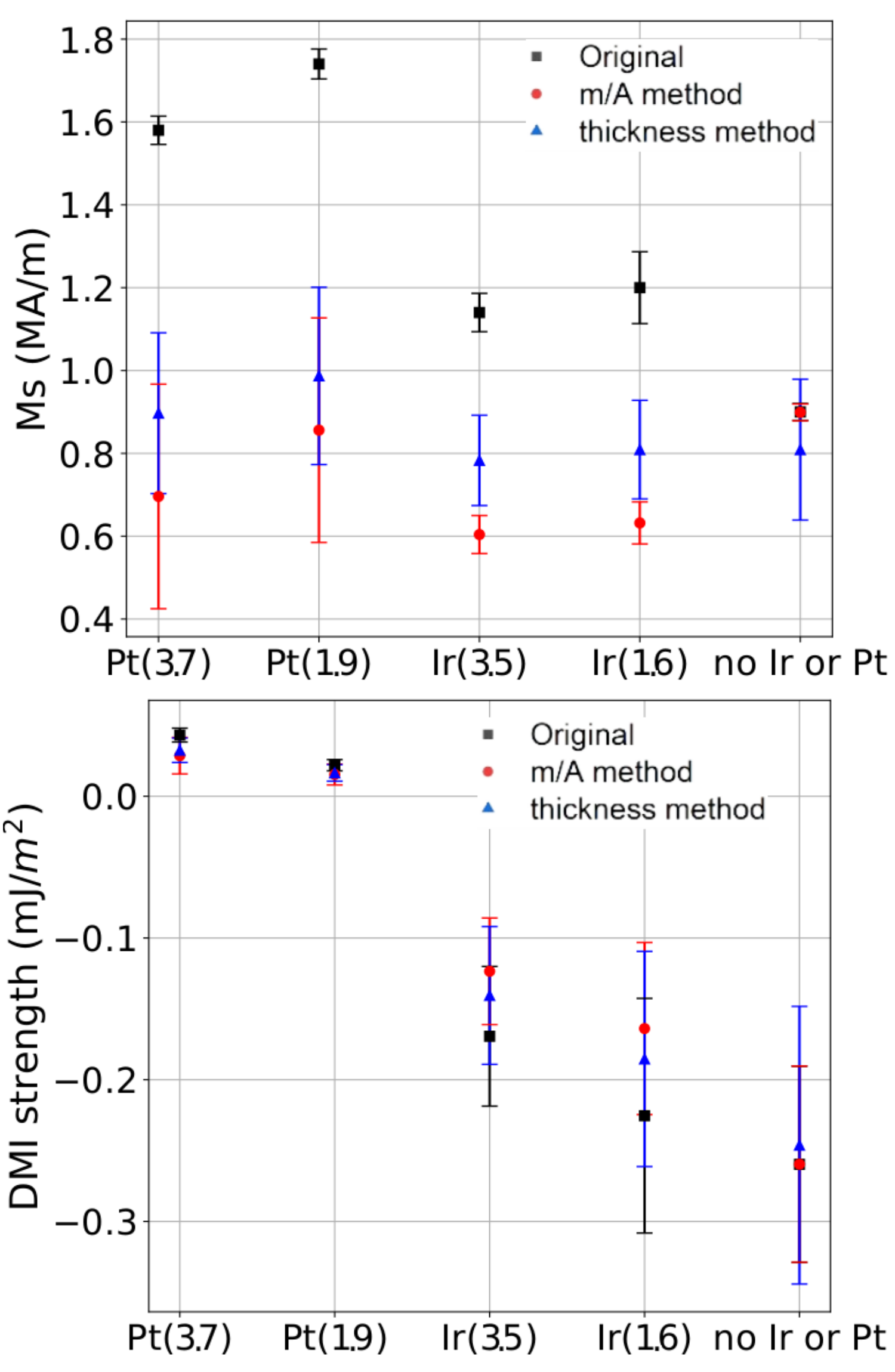}
    \caption{Effect of dead or proximity layers on the determination of the saturation magnetization and consequences for the evaluation of $D$. Correction for $M_\mathrm{s}$ (upper panel) and $D$ (lower panel) obtained by three different analysis of the SQUID measurements. The correction was applied to the total magnetic moment (moment/area method, red circle) and the magnetic layer thickness (``thickness''' method, blue triangle). ``Original'' (black square) refers to the original method taking into account the measured Co thickness.}
    \label{fig:Mscalc}
\end{figure}

However, the effect on the DMI strength is less pronounced due to the corresponding DMI field values having an opposite trend in magnitude to the difference in the $M_s$ values. Nevertheless, the final values for the DMI strength can vary by up to 34\% due to using an uncorrected $M_\mathrm{s}$ value. The comparison of the corrected $M_\mathrm{s}$ and the effect of the correction on the $D$ value is shown in Fig.~\ref{fig:Mscalc}.

\begin{figure}[t]    
 \centering
    \includegraphics[width=0.8\linewidth]{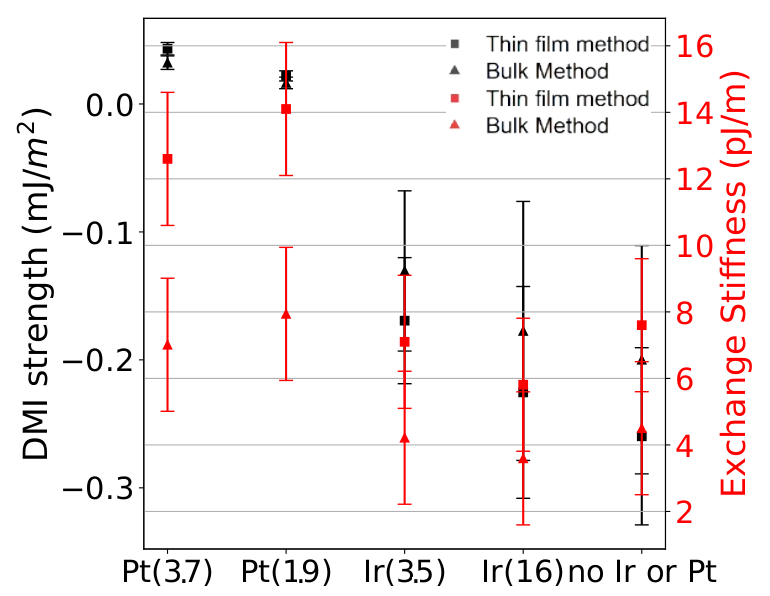}
    \caption{Effect of different methods for the determination of the exchange stiffness on the evaluation of D. The $D$ value (black) and exchange stiffness (red) are shown, calculated by the bulk (triangles) and the thin film (squares) methods (as described in the text).
     }
    \label{fig:Mscalc-D}
\end{figure}

Furthermore, also the exchange stiffness value $A$, which is necessary to calculate the domain wall thickness $\lambda$ (Eq.~\ref{eq:lambdaBloch}) may lead to errors in $D$. It was measured by fitting the saturation magnetization as a function of temperature by the Bloch law $M(T)/M(0)=1-C\frac{4S^2 k_\mathrm{B} T}{A a_0}^{3/2}$  for a thin film \cite{NEM-15}. In this equation, $C$ was assumed to be 0.0294 for fcc lattices, $a_0=$0.355~nm  for cobalt, $S=1$, and $k_\mathrm{B}$ is the Boltzmann constant. Therefore, a further discrepancy in the DMI strength can come from using the bulk material formula to work the exchange stiffness instead of the method for thin films \cite{NEM-15,MOH-19}: in Fig.~\ref{fig:Mscalc-D} the values of $D$ and $A$ are determined by both the bulk and the thin film method (for $M_s$ the original method was used here). The difference in methods causes the DMI strength to vary by 25\%.

\bibliography{dmi_paper}

\end{document}